\documentclass{aa}  

\def \arcsec {\hbox{$^{\prime\prime}$}}

\usepackage{natbib}     
\usepackage[colorlinks=true, citecolor=blue]{hyperref} 
\usepackage{graphicx}
\usepackage{hyperref}
\usepackage{subcaption}
\usepackage{booktabs} 
\usepackage{amsmath} 
\usepackage{float}
\usepackage{multirow}
\usepackage{placeins}
\usepackage{sidecap}
\usepackage{tikz}
\usepackage[switch]{lineno}
\nolinenumbers

\hypersetup{
    colorlinks = true,
    linkcolor = blue,
    citecolor = blue,
    urlcolor = blue
}
\usepackage{multirow}
\usepackage{multicol}
\usepackage{caption}
\usepackage{makecell}
\usepackage{txfonts}
\begin{document} 

   \title{Combined LOFAR-uGMRT analysis of the diffuse radio emission in the massive clusters Abell 773 and Abell 1351}

\author{
K. S. L. Srikanth\inst{1,2}
\and A. Botteon\inst{1}
\and R. Cassano\inst{1}
\and G. Brunetti\inst{1}
\and A. Bonafede\inst{1,2}
\and L. Bruno\inst{1}
\and M. Balboni\inst{2,3}
\and H. Bashir\inst{9}
\and M. Brüggen\inst{4}
\and S. Chatterjee\inst{5}
\and V. Cuciti\inst{1,2}
\and D. Dallacasa\inst{1,2}
\and A.Datta\inst{11}
\and F. de Gasperin\inst{1}
\and G. Di Gennaro\inst{1}
\and C. Groeneveld\inst{1,6}
\and R. Kale\inst{8}
\and M. A. Malik\inst{9}
\and S. Paul\inst{10}
\and S. Salunkhe\inst{8}
\and R. J. van Weeren\inst{6}
\and T. Venturi\inst{1,5}
\and X. Zhang\inst{7}
}

\institute{
Istituto Nazionale di Astrofisica (INAF) - Istituto di Radioastronomia (IRA), via Gobetti 101, 40129 Bologna, Italy
\and Dipartimento di Fisica e Astronomia (DIFA), Università di Bologna, via Gobetti 93/2, 40129 Bologna, Italy
\and Istituto Nazionale di Astrofisica – Istituto di Astrofisica Spaziale e Fisica Cosmica (IASF), Via A. Corti 12, 20133 Milano, Italy
\and Hamburger Sternwarte, Universität Hamburg, Gojenbergsweg 112, 21029 Hamburg, Germany
\and Center for Radio Astronomy Techniques and Technologies, Rhodes University, Grahamstown 6140, South Africa
\and Leiden Observatory, Leiden University, PO Box 9513, 2300 RA Leiden, The Netherlands
\and Max-Planck-Institut für Extraterrestrische Physik (MPE), Gießenbachstraße 1, D-85748 Garching bei München, Germany
\and National Centre for Radio Astrophysics, Tata Institute of Fundamental Research, S. P. Pune University Campus, Ganeshkhind, Pune 411007, India
\and University of Kashmir, Hazratbal, Srinagar, J\&K, India – 190006
\and Manipal Centre for Natural Sciences, Manipal Academy of Higher Education, Karnataka, Manipal - 576104, India
\and Indian Institute of Technology Indore, Indore-453552, M.P., India.\\
\email{koushikasri.srikant2@unibo.it}
}

   \date{January 15, 2026 }

  \titlerunning{LOFAR-uGMRT study of A773 and A1351}
 \authorrunning{Srikanth et al.}
  \abstract
   {Radio halos are megaparsec-scale diffuse, non-thermal radio sources located at the center of galaxy clusters. They trace relativistic particles and magnetic fields in the intra-cluster medium. The source of energy for their formation is believed to be the merging of galaxy clusters, which generates turbulence and re-accelerates aged electrons.}
   {We aim to study the diffuse radio emission, spectral properties and connection between thermal and non- thermal emission in the massive ($M_{500} \sim 7 \times 10^{14} M_{\odot}$), dynamically disturbed galaxy clusters Abell 773 and Abell 1351.}
   {We combine observations from the LOFAR Two-meter Sky Survey - Data Release 2 (LoTSS-DR2) at 144 MHz and the new upgraded Giant Meterwave Radio Telescope (uGMRT) at 650 MHz for both clusters. Additionally, archival \textit{XMM-Newton} X-ray images are utilized to supplement our analysis.}
   {We confirm that both clusters host a radio halo, extending up to a largest linear size (LLS) of $\sim$ 2 Mpc. 
   We measure an integrated spectral index \(\alpha_{144}^{650}\) of $\sim$ -1.0 for both the clusters. 
   We show that radio halo in A773 resembles a classical radio halo that follows a sublinear relation in point-to-point analysis between radio and X-ray surface brightness. 
Conversely, A1351 exhibits a more complex and asymmetric radio halo, embedded by several radio sources, including the bright cluster galaxy (BCG), a tail galaxy (TG), and the ridge. We found a deviation from the sublinear relation in the point-to-point analysis of A1351 due to the presence of these contaminating radio sources.}
   {}


   \maketitle
%
\linenumbers
\section{Introduction}
Galaxy clusters are 
gravitationally bound structures with a mass \(\sim\) $10^{14} - 10^{15}$ $M_{\odot}$. According to the Lambda Cold Dark Matter ($\Lambda$CDM) model, clusters form  hierarchically through sequence of mergers and accretion of smaller systems over a timescale of about $10^{9} \sim 10^{10}$ years \citep{press74, springel06}. The space between the galaxies is filled by the hot intra-cluster medium (ICM), a plasma with a temperature of  \(\sim\) $ 10^{7}-10^{8} \mathrm{K}$. Mergers between galaxy clusters are the most energetic events in the Universe. While most of this energy contributes to the heating of the ICM, part of it is also channelled into the acceleration of relativistic particles via shocks and turbulence and amplification of magnetic fields \citep{markevitch07rev}.

Radio observations of galaxy clusters have uncovered large-scale, diffuse synchrotron sources that trace the presence of relativistic particles and magnetic fields in the ICM. These sources are classified as radio relics and radio halos, respectively located at the peripheries and centers of merging clusters, and mini-halos surrounding the cool core region of the relaxed clusters \citep{vanweeren19rev}. 
Among these, radio haloes (RHs)
often extend up to $\sim$ Mpc scales and are likely powered by turbulent re-acceleration process within the ICM \citep{brunetti14rev}. An alternative explanation is that radio halos form as synchrotron emission by secondary electrons generated by proton-proton collisions in the ICM \citep{dennison80,blasi99,dolag00}. 
However, the lack of \(\gamma\)-ray detections from galaxy cluster observations expected as by-product of the proton-proton collisions \citep{brunetti17,adam21} and the existence of radio halos with very steep radio spectra ($\alpha<1$, where $f(\nu) \propto \nu^{\alpha}$) \citep{brunetti08,wilber18a1132,digennaro21highz,bruno21,pasini2024ultra, osinga2024probing,santra2024deep} 
disfavor hadronic models as the primary mechanism for their origin. This is further supported by the sublinear point-to-point correlations observed between the radio and X-ray surface brightness in clusters hosting radio halos \citep{govoni01comparison,brunetti14rev}. Such correlations suggest that the non-thermal components exhibit a weak radial decline with respect to the thermal gas \citep{balboni2024chex}. 
In re-acceleration models, the energy available for turbulent re-acceleration comes from the dissipation of gravitational energy of mergers, and hence a correlation between the occurrence of radio halos and the cluster mass and dynamics is expected \citep{cassano05,cassano10mc,wen13,giacintucci17}. 
Furthermore, several studies have revealed that the majority of RHs are found in merging clusters, whereas clusters without diffuse emission are typically in a relaxed state 
\citep{cassano10connection, kale15, venturi07, venturi08,cassano13,cassano2023planck}. 

A key prediction of turbulent re-acceleration models is that a
large fraction of halos should have very steep spectra, being generated during less energetic merger events (i.e.; minor mergers or major mergers between less massive clusters) where the turbulent energy may not be enough to accelerate electrons up to the energy required to generate synchrotron radiation at GHz frequencies \citep{cassano06,brunetti08}. Recent statistical studies at low frequency with LOFAR have indeed identified RHs also in less disturbed  clusters \citep{cassano2023planck}. Additionally, there is evidence of cluster-scale diffuse emission in cool-core clusters \citep{bonafede14cl, savini18planck139, biava2024first, van2024lofar}.
In this paper, we focus on the study of two massive and optically rich galaxy clusters, PSZ2~G166.09+43.38 (also known as A773) and PSZ2~G139.18+56.37 (also known as A1351), both hosting giant RH. The clusters were observed using the LOw Frequency ARray High Band Antenna (LOFAR-HBA; 120--168 MHz) and the upgraded Giant Metrewave Radio Telescope Band 4 (uGMRT Band 4; 550--750 MHz). The physical properties of the clusters are listed in Table~\ref{tab:Properties of galaxy clusters}. From X-ray analysis both A773 and A1351 are identified as merging clusters but in different merger states. Chandra X-ray observations reveal that A773 is an advanced merger \citep{barrena2007internal}. A1351, on the other hand, is an ongoing merger \citep{allen2003cosmological, giovannini09, giacintucci09a1351, barrena14}, a scenario further supported by weak-lensing analyses from \citet{holhjem09}.
The presence of a giant RH in A773 ($\sim 1.6$ Mpc) was first reported by \cite{govoni01six} using Very Large Array (VLA) observations at 1.4 GHz in C and D configurations. The diffuse radio emission in A1351 was initially detected by \cite{owen99rass}, 
and later VLA observations at 1.4 GHz by \cite{giacintucci09a1351} confirmed a giant RH with an extent of $\sim 1.1$ Mpc. More recently, \cite{chatterjee2022unveiling} used GMRT (610 MHz) and VLA (1.4 GHz) observations to provide additional insights into the RH structure and the radio ridge in A1351.

\begin{table*}[ht]
\caption{Galaxy cluster properties.}
\label{table:cluster_properties}
\centering
\begin{tabular}{p{3.5cm} p{6.2cm} p{6.2cm}} 
\toprule
\textbf{Properties} & \textbf{PSZ2 G166.09+43.38 (A773)} & \textbf{PSZ2 G139.18+56.37 (A1351)} \\
\midrule
Optical Richness & Optically rich, Abell class 2 \tablefootmark{(a)} & Optically rich, Abell class 2\tablefootmark{(a)} \\
Redshift (\(z\)) & 0.217 & 0.322 \\
\(M_{500}\) (\(M_\odot\)) & \(6.85 \times 10^{14}\)\tablefootmark{(d)} & \(6.87 \times 10^{14}\)\tablefootmark{(d)} \\
X-ray Luminosity (\(L_X\)) & \(12.5 \times 10^{44}\, h_{50}^{-2}\, \text{erg\,s}^{-1}\)\tablefootmark{(e)} & \(8.4 \times 10^{44}\, h_{50}^{-2}\, \text{erg\,s}^{-1}\)\tablefootmark{(f)} \\
ICM Temperature (\(T_X\)) & \(7\text{--}9\,\text{keV}\)\tablefootmark{(g)} & \(\sim 9\,\text{keV}\)\tablefootmark{(c)} \\
Concentration parameter (\(c\)) & \(1.84 \times 10^{-1} \pm 6.80 \times 10^{-3}\)\tablefootmark{(h)}
 &  \(8.6 \times 10^{-1} \pm 5.9 \times 10^{-3}\)\tablefootmark{(h)}\\
Centroid shift (\(w\)) & \(1.83 \times 10^{-1} \pm 5.71 \times 10^{-3}\)\tablefootmark{(h)}
 &  \(4.7 \times 10^{-1} \pm 8.49 \times 10^{-3}\)\tablefootmark{(h)}\\
\bottomrule
\label{tab:Properties of galaxy clusters}
\end{tabular}
\tablefoot{
\tablefoottext{a}{\citet{dahle02, abell89}.},
\tablefoottext{b}{\citet{barrena2007internal}.},
\tablefoottext{c}{\citet{barrena14}.},
\tablefoottext{d}{\citet{planck16xxvii}.},
\tablefoottext{e}{\citet{ebeling96}.},
\tablefoottext{f}{\citet{allen2003cosmological, bohringer00}.},
\tablefoottext{g}{\citet{rizza98, govoni04chandra, ota04},}
\tablefoottext{h}{\citet{botteon22, zhang2023}.}
}
\end{table*}

This paper is structured as follows: In Section~\ref{sec:Radio observation and data analysis}, we describe the radio observations, data reduction, and analysis. Section~\ref{sec:cluster intro} presents the results obtained from the radio observations. A discussion of these results is provided in Section~\ref{sec:Discussion}, followed by a summary and conclusions in Section~\ref{sec:conclusion}.

Throughout this work, we adopt a standard $\Lambda$CDM cosmology with $H_0 = 70\,\mathrm{km\,s^{-1}\,Mpc^{-1}}$, $\Omega_\mathrm{M} = 0.3$, and $\Omega_\Lambda = 0.70$.

\section{Radio observation and data analysis}\label{sec:Radio observation and data analysis}
\subsection{LOFAR observations}\label{subsec: LOFAR observations}
The clusters A773 $\&$ A1351 were observed as part of the LoTSS-DR2 survey \citep{shimwell22}\footnote{Both clusters belong to the sample of PSZ2 clusters detected in the LoTSS-DR2 \href{https://lofar-surveys.org/planck_dr2.html}{(LoTSS-DR2/PSZ2)}.} which covers 27\% of the northern sky in the frequency range of 120--168 MHz (central frequency 144 MHz). Each LoTSS pointing lasts 8 hours, achieving a median root-mean-square ($\sigma_{rms}$) sensitivity of $\sim 0.1 \, \mathrm{mJy \, beam^{-1}}$ at a resolution of $6^{\prime\prime}$ \citep{shimwell19}. We utilized the data processed by \cite{botteon22} in their study to search for radio halos in the Planck clusters within the LoTSS-DR2 survey. For detailed information on the data processing of the clusters in LoTSS-DR2, we refer the reader to \citet{shimwell19,shimwell22} and \citet{tasse21}, which describe the Survey Key Project (SKP) pipelines. Direction-independent and direction-dependent calibration and imaging were performed using \textsc{PREFACTOR} \citep{vanweeren16calibration, williams16, degasperin19} and the \texttt{ddf-pipeline}, which includes \textsc{DDFacet} \citep{tasse18} and \textsc{KILLms} \citep{tasse14arx, tasse14, smirnov15}. The quality towards the targets was improved by \citet{botteon22} using the extraction and self-calibration procedure described in \citet{vanweeren21}. We re-imaged the extracted and self-calibrated data instead of using the images from \cite{botteon22} to perform a tailored subtraction of discrete sources corresponding to our clusters. The source-subtracted images in \cite{botteon22} were produced with an inner uv-cut of \(80\lambda\) to eliminate large-scale Galactic emission (\(\sim 40'\)), while here we applied a higher inner uv-cut of \(200\lambda\) (corresponding to physical scales of 3.6 Mpc for A773 and 4.5 Mpc for A1351) 
during imaging, matching the shortest baseline of the uGMRT, to ensure consistency between the images at both frequencies. Imaging was performed using \textsc{WSClean v3.6} \citep{offringa14} with Briggs weighting \citep{briggs95} using a robust parameter of $-0.5$ and applying different \textit{uv}-tapers to produce images at varying resolutions. The multi-scale, multi-frequency deconvolution algorithm \citep{offringa17} was enabled 
and the bandwidth was divided into six frequency channels per imaging run. 
The systematic uncertainties due to flux density scale calibration is set to 10$\%$ for LoTSS-DR2 \citep{shimwell22}.

\subsection{uGMRT observations}\label{subsec:uGMRT observations}
Both clusters were observed with the uGMRT in GMRT Software Backend (GSB) and GMRT Wideband Backend (GWB) Band 4. GSB has a central frequency of 610 MHz with a bandwidth of 32 MHz subdivided into 256 channels. For GWB, the central frequency was 650 MHz, observed with a bandwidth of 200 MHz split into 2048 channels. Both GSB and GWB had an integration time of 4s and measured only total intensity. (Project code: 43\_023, P.I. R.Cassano). Each cluster was observed in 6 hr parallel in GSB and GWB. 3C48 was used as the primary calibrator for the processing of A773, and 3C386 was used for A1351. Data were processed using \texttt{Source Peeling and Atmospheric Modeling \citep[SPAM][]{intema09_soft}}.
Using the final GSB images, a source catalog was generated with \texttt{Python Blob Detector and Source Finder \citep[PyBDSF][]{mohan15}} to model the sky and perform direction-dependent calibration on GWB. GWB data were split into four 50 \,MHz sub-bands for calibration. Finally, the calibrated sub-bands were imaged using the same procedure as that of LOFAR imaging except that the bandwidth was divided into four frequency channels per imaging run (i.e, 50MHz per channel). Similar to LOFAR imaging, an inner \textit{uv}-cut of 200$\lambda$ was applied to the images. The systematic uncertainties due to flux scale calibration are set to 5$\%$ for the observations in Band 4 \citep{chandra04}.  
\begin{table*}[h]
    \centering
    \small
    \resizebox{\textwidth}{!}{%
    \begin{tabular}{|l|c|c|c|c|c|c|}
        \hline
        \textbf{Cluster name} & \textbf{Central frequency [MHz]} & \textbf{Resolution [$\arcsec \times \arcsec$]} & \textbf{Beam position angle [deg]} & \textbf{uv min [$\lambda$]} & \textbf{uv-taper [$\arcsec$]} & \textbf{$\sigma_\mathrm{rms}$ [$\mu$Jy beam\textsuperscript{-1}]} \\
        \hline
        \multirow{4}{*}{PSZ2 G166.09+43.38 (A773)} & 144 & 7.9$\times$4.6 & 92.3 & 200 & -- & 76.0 \\
        & 144 & 34.2$\times$31.6 & 7.9 & 200 & 28.4 & 203.9 \\
        & 650 & 8.8$\times$2.9 & 75.5 & 200 & -- & 16.3 \\
        & 650 & 31.6$\times$16.6 & 54.7 & 200 & 20 & 37.7 \\
        \hline
        \multirow{4}{*}{PSZ2 G139.18+56.37 (A1351)} & 144 & 8.6$\times$4.3 & 97.8 & 200 & -- & 67.2 \\
        & 144 & 28.0$\times$24.1 & 30.2 & 200 & 21.3 & 163.3 \\
        & 650 & 4.6$\times$3.3 & 150.1 & 200 & -- & 12.5 \\
        & 650 & 23.8$\times$17.8 & 173.2 & 200 & 20 & 57.4 \\
        \hline
    \end{tabular}%
    }
    \caption{Image properties of the clusters A773 \& A1351.}
    \label{table:clusters_rms}
\end{table*}

\subsection{Integrated synchrotron spectra and source subtraction}\label{subsec:integrated spectra}
To optimize the analysis of diffuse emission, images were generated using various gaussian tapers to increase the weighting of shorter baselines for both LOFAR and uGMRT (see Table. ~\ref{table:clusters_rms}). The noise levels of the high-resolution and low-resolution images of
LOFAR and uGMRT are reported in Table~\ref{table:clusters_rms}. Since the focus of this study is the radio halo emission, compact sources or contaminants within or near the halo emission were subtracted out. Initially, we created a high-resolution image containing only the compact sources with a Briggs parameter of $-0.5$ and a \textit{uv}-cut corresponding to a physical scale of 250~kpc  (consistent with \cite{botteon22}) at the cluster redshift (A773 - 2893$\lambda$, A1351 - 3892$\lambda $) to remove their contribution from large scale diffuse emissions. The model visibilities of the discrete sources were obtained from \textsc{WSclean} \texttt{-predict} function and they were subtracted from the original data. The subtracted visibilities were subsequently imaged enabling the \texttt{-multi-scale} de-convolution algorithm for different taper values (5\arcsec, 10\arcsec, 15\arcsec and 20\arcsec). A mask was generated from this image by selecting clean components above a $3\sigma$ ($\sigma$ is the rms noise of the image) threshold and then applied in a final imaging run using \texttt{-fits-mask} and with an inner uvcut of 200$\lambda$. Following it, the LOFAR and uGMRT images were convolved to a common resolution using \textsc{CASA} to enable a consistent comparative analysis. The image in which maximum extended emission was recovered ($50 \arcsec \times 50 \arcsec$ for A773 and $33\arcsec \times 33\arcsec$ for A1351 Fig. \ref{fig:A773 Low resolution}, \ref{fig:A1351 Low resolution}) without compromising the image fidelity was used to find the flux density of the halos. 
To test the consistency of this source subtraction process, we compared the flux density obtained from the source-subtracted images with that derived by algebraically subtracting the flux densities of the discrete sources from the total flux of the halo (including discrete sources). The results were consistent within $1\sigma$ uncertainty for both the LOFAR and uGMRT images.

\subsection{Flux density and spectral index measurement}\label{sec:Flux density and spectral index measurement}
To measure the flux density of the radio halo in LOFAR and uGMRT images, we adopted two methods which are explained in the next section. 
\subsubsection{2 $\sigma$ contours}\label{subsec: 2 sigma contours}
In the first method, we integrated the flux density within the region defined by the $2\sigma$ contours of the radio halo. To assess consistency and potential differences in the emission size between LOFAR and uGMRT, we used two regions: one defined by the LOFAR $2\sigma$ contours and the other by the uGMRT $2\sigma$ contours. For each of these regions, we measured the integrated flux density in both the LOFAR and uGMRT images and derived the corresponding spectral slope. The uncertainity in the flux density was calculated by the following formula,
\begin{equation}
\Delta S_{\nu} = \sqrt{(f S_{\nu})^2 + N_{\text{beam}} \sigma_{\text{rms}}^2 + \sigma_{\text{sub}}^2} \,,
\end{equation}
where \( f \) represents the systematic uncertainties due to absolute flux scale errors (see Sec.\ref{subsec: LOFAR observations}, \ref{subsec:uGMRT observations}). 
\( {\sigma_{rms}} \) is the map noise level, \( N_{\text{beam}} \) is the number of beams covering the halo region, and \( \sigma_{\text{sub}} \) is the uncertainty of the source subtraction in the \( uv \)-plane which was calculated by following the procedure described in \cite{botteon22}. 
The flux density values are listed in the Table \ref{tab:flux and slopes}.\\
From the measured flux densities, we determined the spectral index $\alpha$ and its error. 
The error in the $\alpha$ was determined by
\begin{equation}
    \Delta \alpha = \frac{1}{\ln \frac{\nu_1}{\nu_2}} \sqrt{\left( \frac{\Delta S_1}{S_1} \right)^2 + \left( \frac{\Delta S_2}{S_2} \right)^2 }.
    \label{eq:equation1}
\end{equation}

\subsubsection{LOFAR model injection}\label{subsec: Lofar model injection}
Injection of an exponential model has been the most widely employed technique to estimate upper limits in cases of radio halo non-detection, often caused by missing short baselines or poor sensitivity \citep{brunetti07cr, venturi08, kale13, bruno2023planck}. As a step forward to this approach, we adopt a method with a key improvement of using a realistic model: using LOFAR model obtained during the LOFAR imaging process, rather than an idealized exponential profile to inject into the uGMRT visibilities. 
A similar strategy was employed by \cite{giacintucci14rxj1720}, who injected the CLEAN components of a mini-halo observed with the GMRT at 617 MHz into Very Large Array (VLA) data at 4.86 GHz and 8.44 GHz for different spectral indices. This method not only provides robust upper limits, but also allows us to quantify the flux density loss attributed due to instrumental limitations. It offer us a more realistic estimate of the halo flux and the spectral index and an opportunity to check for possible systematics in the spectral index estimate. 

The following steps briefly describe the injection method:

We begin the analysis by assuming a flux density to be injected into the uGMRT visibilities. The LOFAR model image is then rescaled to this flux density by adopting the spectral index $\alpha$ required to reproduce the corresponding value at 650 MHz. The rescaling is applied to the CLEAN components associated with the radio halo in the LOFAR low-resolution model, using an inner \textit{uv}-cut of 80\(\lambda\), to include all recoverable diffuse emission by LOFAR. The offset to inject the radio halo is chosen to be close enough to the center to avoid significant primary beam attenuation, while ensuring that the region is free of contaminating sources or imaging artifacts (see Fig.~\ref{fig:injected_images_sec2}). 
After the injection, the "updated" uGMRT dataset is imaged following the standard procedure (see Sect. \ref{subsec:uGMRT observations}, Sect. \ref{subsec:integrated spectra}), and the recovered flux density of the injected halo is measured within the 2$\sigma$ contour of the new image. To account for local noise fluctuations and and potential contributions from discrete sources, the flux density measured in the same region prior to injection is subtracted from the recovered flux.
If the recovered flux density lies within $\pm 5\%$ of the flux density measured in the observations within 2$\sigma$ contours, the corresponding injected flux density is considered a reliable estimate of the halo’s flux density.  Consequently, the spectral index $\alpha$ used to obtain this injected flux is regarded as the reliable spectral index of the halo. This value should be considered a lower limit on \(\alpha\), as it assumes no flux density loss in LOFAR. \citet{bruno2023planck} (see their fig.~14) showed that flux density losses in LOFAR observations are $\le$ 10\% for radio halos with diameters between 7.5$'$ and 10.5$'$. However, over the broad frequency range of 144-650 MHz, such flux density losses have a negligible impact on the spectral index $\alpha$. Since the halos considered in this study have maximum sizes of $\sim$10$'$, 
LOFAR flux density losses are expected to be minimal, and the model images are deemed suitable for injection. The procedure is then repeated iteratively using different values of injection fluxes until the recovered flux matches the observed flux. 
If the spectral slope derived from observations between the LOFAR and uGMRT images is steeper than the slope obtained by the injection method, it indicates that the uGMRT has suffered losses due to missing baselines and sensitivity issues. 

\section{Results}\label{sec:cluster intro}
\begin{figure*}[h]
    \centering
    \begin{subfigure}[b]{0.48\textwidth}
        \centering
        \includegraphics[width=9cm, height=7cm]{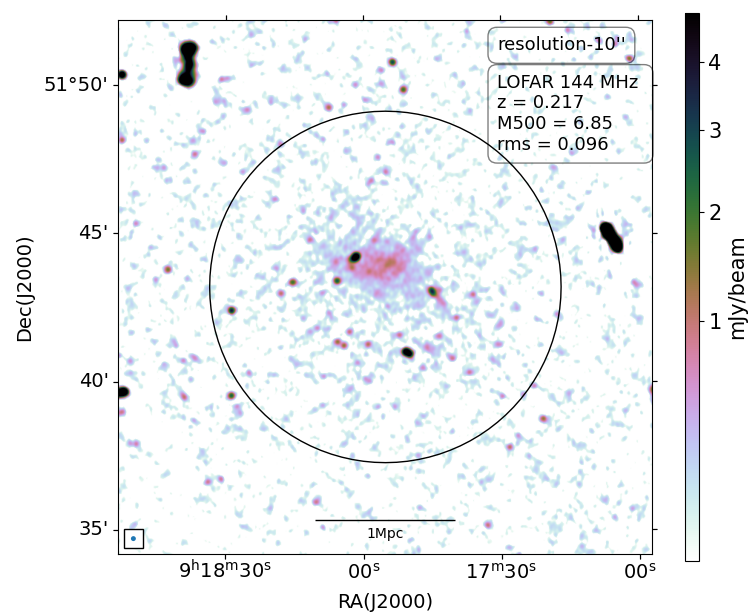}
        \label{fig:A773 lofar con10}
    \end{subfigure}
    \hfill
    \begin{subfigure}[b]{0.48\textwidth}
        \centering        \includegraphics[width=9cm, height=7cm]{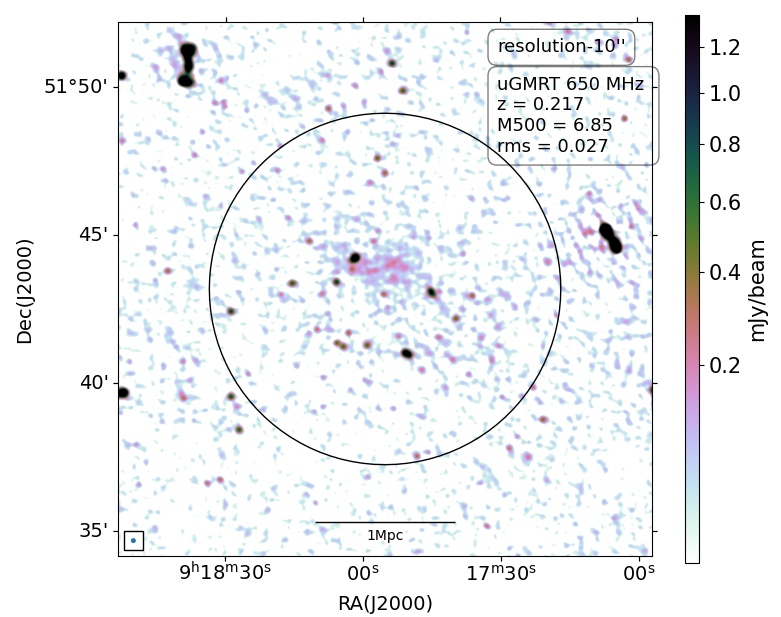}
        \label{fig:A773 gmrt con10}
    \end{subfigure}
    \caption{\textit{Left:} High resolution LOFAR image of A773 at 144 MHz. \textit{Right:} High resolution uGMRT image of A773 at 650 MHz. Both images are at a resolution of \(10'' \times 10''\), with \(M_{500}\) in units of \(10^{14} M_\odot\) and rms in mJy/beam. The drawn region represents the \(r_{500}\) scale of the cluster.}
    \label{fig:A773 high resolution}
\end{figure*}

High-resolution LOFAR and uGMRT images of A773 are reported in Fig. \ref{fig:A773 high resolution}.
The central regions of the cluster show the presence of multiple radio sources. 
At the resolution of \(10\arcsec\), the radio halo detected in LOFAR is larger in size than uGMRT (Fig \ref{fig:A773 high resolution}, left panel).
\begin{figure*}[h]
    \centering    \includegraphics[width=0.93\textwidth]{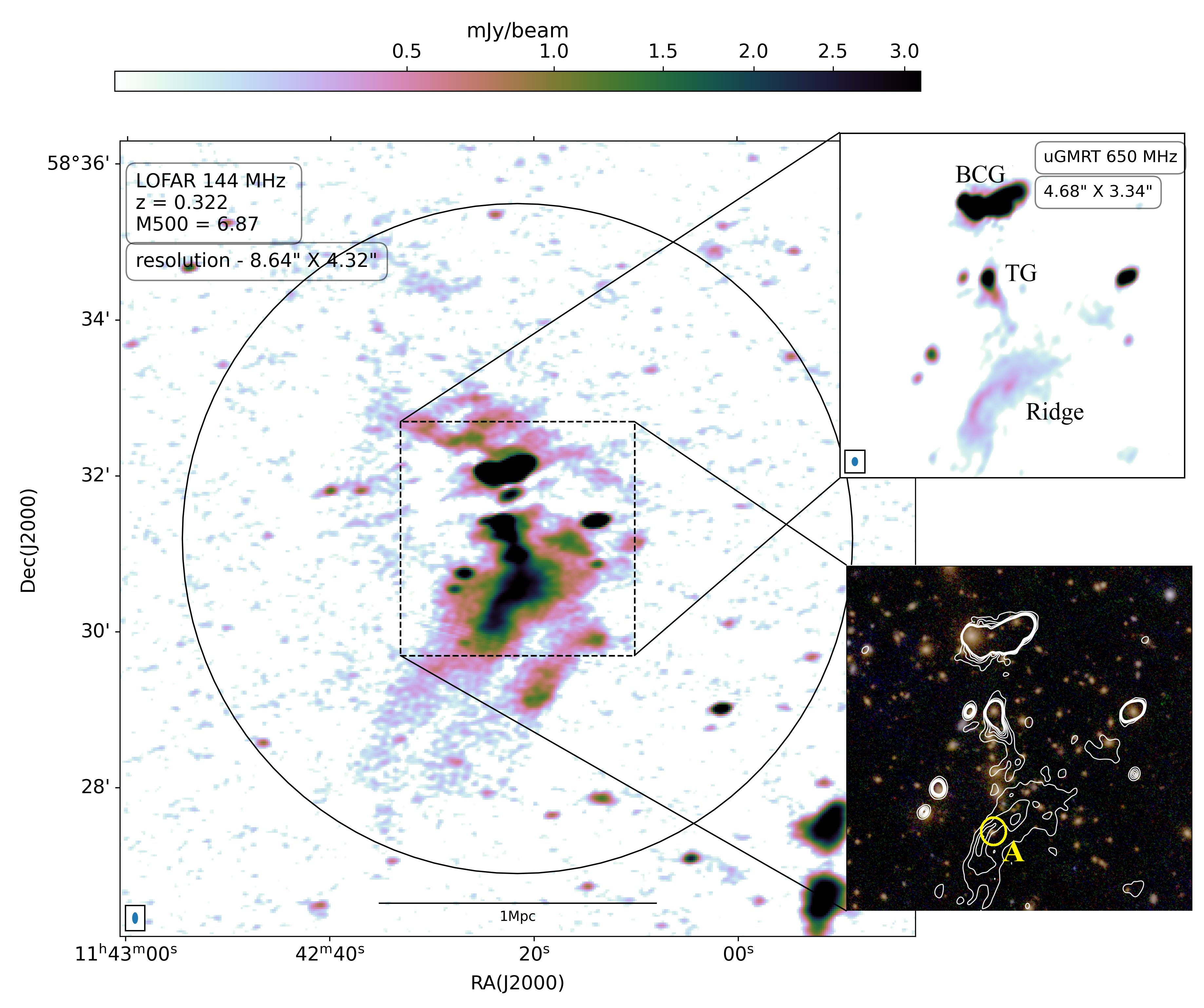}
    \caption{High-resolution LOFAR image of A1351 at 144\,MHz with a resolution of \(8.64'' \times 4.32''\) and an rms noise of 68\,$\mu$Jy\,beam$^{-1}$. The black circle represents the \(r_{500}\) scale of the cluster. Inset top: Zoom-in of the cluster center at uGMRT 650\,MHz  at a resolution of \(4.68'' \times 3.34''\). The rms of the uGMRT image is 13\,$\mu$Jy\,beam$^{-1}$. Inset bottom: Optical Pan-STARRS (g,r,i) image of the cluster center overlaid with uGMRT radio contours at levels of (9, 18, 27, 36, 45, 52) $\times  ~\sigma_{rms}$.  \(M_{500}\) is given in units of \(10^{14}\,M_\odot\).}
    \label{fig:A1351_high_resolution_zoomed}
\end{figure*}\\
High-resolution images of A1351 (see Fig. \ref{fig:A1351_high_resolution_zoomed}) show a more complex radio morphology which include different sources, including one of the two brightest cluster galaxies (BCG), a tailed galaxy (TG), and an extended region of diffuse emission referred to as the "ridge" (Fig. \ref{fig:A1351_high_resolution_zoomed}). 
In the LOFAR  high-resolution image, 
we reached a resolution of \(8.64'' \times 4.32''\). However, at this resolution, TG and the ridge are still not resolved. The improved sensitivity and high angular resolution of the uGMRT image (\(4.68'' \times 3.34''\)),  
clearly reveals both the TG and the ridge structure. Notably, the morphological details of the ridge are distinctly resolved. The ridge extends for a projected length of \(\sim 470\) kpc and is located at a projected distance of \(\sim 500\) kpc from the center of the cluster. 
The cluster center was set at the location of BCG, as described in \cite{barrena14}. From the head of the TG, the center of the ridge is \(\sim 330\) kpc away. 
 
\subsection{Radio halo}
\begin{figure*}[h]
    \centering
    \begin{subfigure}[b]{0.33\textwidth}
        \centering
        \includegraphics[width=6.2cm,height=5.2cm]{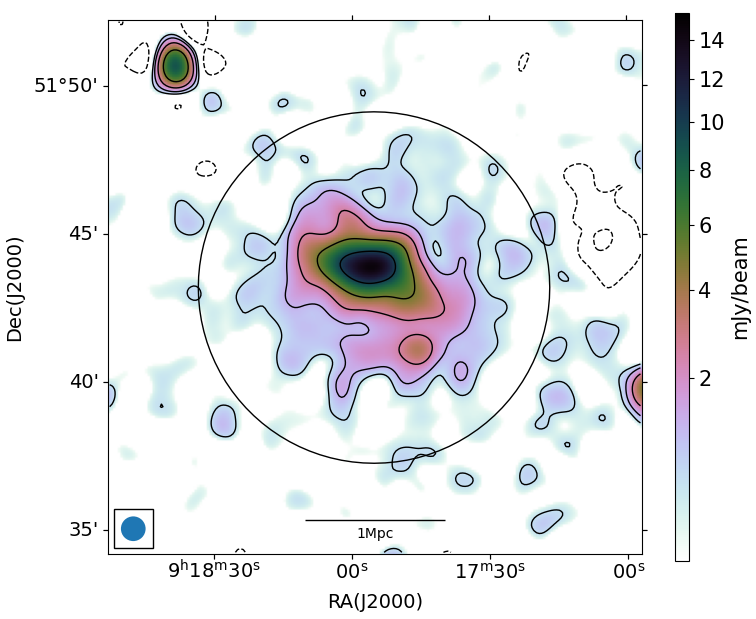}
        \label{fig:A773 lofar con50}
    \end{subfigure}
    \hfill
    \begin{subfigure}[b]{0.33\textwidth}
        \centering
        \includegraphics[width=6.2cm,height=5.2cm]{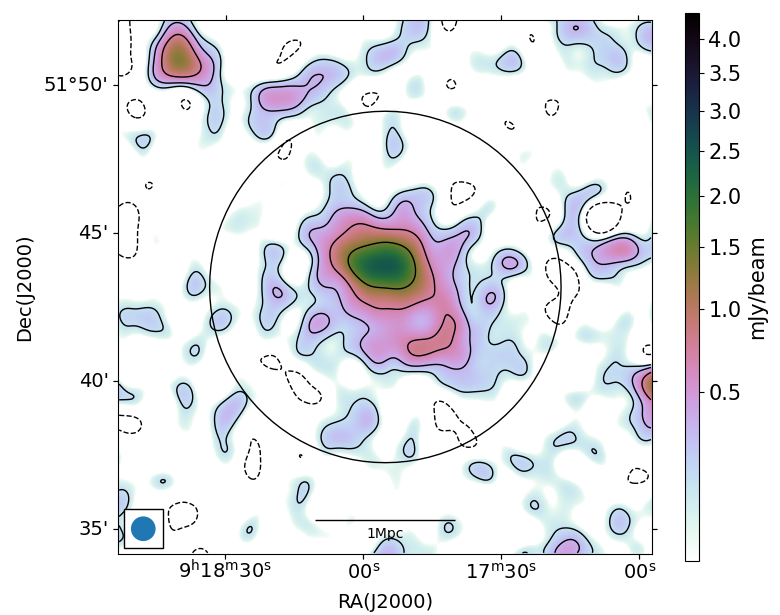}
        \label{fig:A773 gmrt con50}
    \end{subfigure}
    \begin{subfigure}[b]{0.33\textwidth}
        \centering
        \includegraphics[width=6.2cm,height=5.25cm]{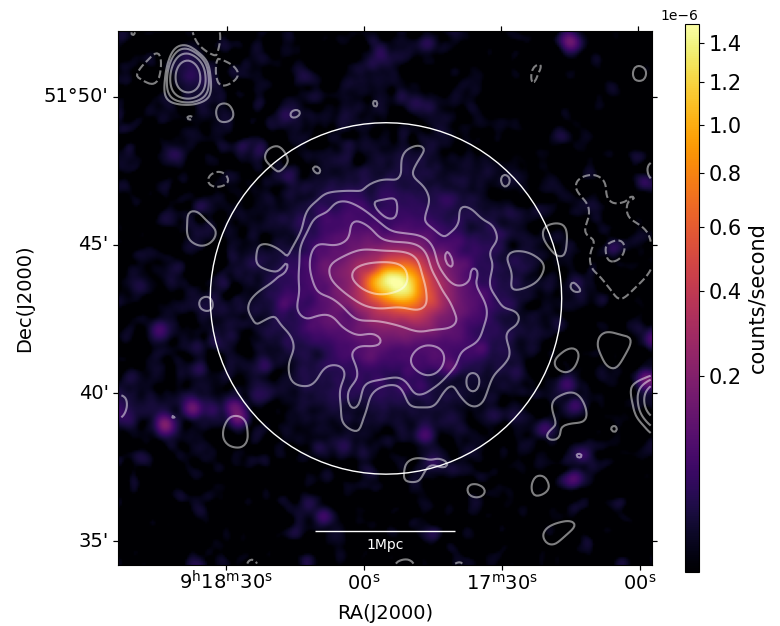}
        \label{fig:A773 XMM}
    \end{subfigure}
    \caption{\textit{Left:} Low resolution source subtracted LOFAR image of A773 at 144 MHz overlaid with LOFAR 2$\sigma$ contours. \textit{Middle:} Low resolution source subtracted uGMRT image of A773 at 650 MHz overlaid with uGMRT 2$\sigma$ contours. \textit{Right:}  XMM image of A773 overlaid with LOFAR source subtracted contours. The contour levels are spaced as $(-2, 2, 4, 8, 16, 32) \times \sigma_{rms}$. LOFAR and uGMRT source subtracted images have a  resolution of \(50'' \times 50''\). The region drawn around the halo represents the \(r_{500}\) scale of the cluster.}
    \label{fig:A773 Low resolution}
\end{figure*}
\begin{figure*}[h]
    \centering
    \begin{subfigure}[b]{0.33\textwidth}
        \centering
        \includegraphics[width=6.2cm,height=5cm]{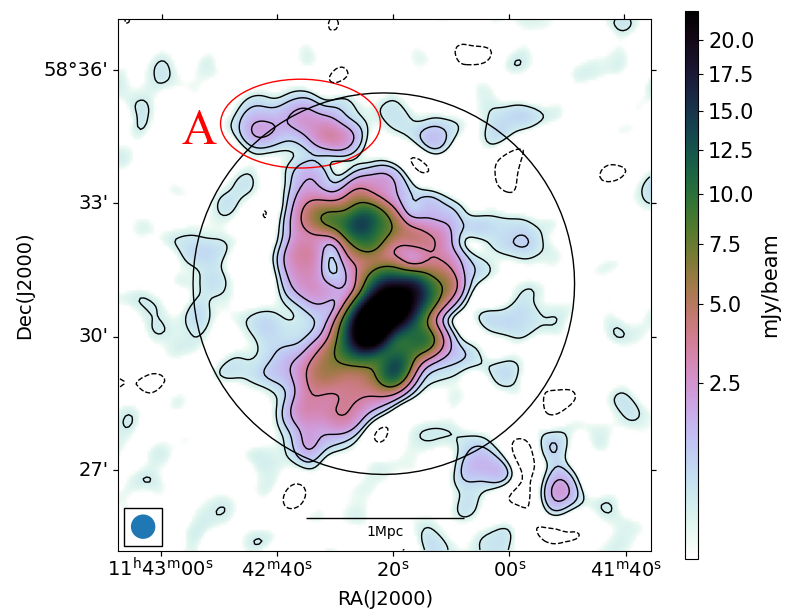}
        \label{fig:A1351 lofar con33}
    \end{subfigure}
    \hfill
    \begin{subfigure}[b]{0.33\textwidth}
        \centering
        \includegraphics[width=6cm,height=5.1cm]{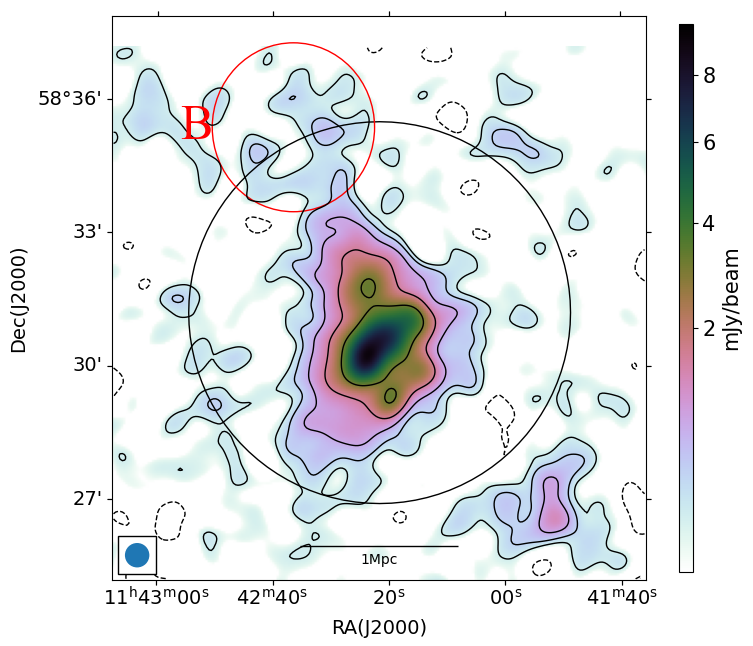}
        \label{fig:A1351 gmrt con33}
    \end{subfigure}
    \begin{subfigure}[b]{0.33\textwidth}
        \centering
        \includegraphics[width=6cm,height=5.2cm]{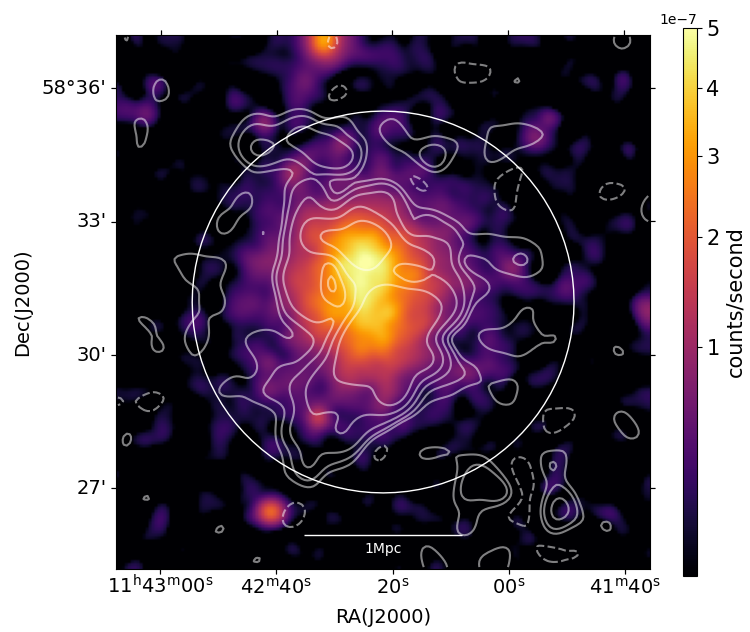}
        \label{fig:A1351 XMM}
    \end{subfigure}
    \caption{\textit{Left:} Low resolution source subtracted LOFAR image of A1351 at 144 MHz overlaid with LOFAR 2$\sigma$ contours. \textit{Middle:} Low resolution source subtracted uGMRT image of A1351 at 650 MHz overlaid with uGMRT 2$\sigma$ contours. \textit{Right:}  XMM image of A1351 overlaid with LOFAR source subtracted contours. The contour levels are spaced as $(-2, 2, 4, 8, 16, 32) \times \sigma_{rms}$. LOFAR and uGMRT source subtracted images have a  resolution of \(33'' \times 33''\). The region drawn around the halo represents the \(r_{500}\) scale of the cluster.}
    \label{fig:A1351 Low resolution}
\end{figure*}
\subsubsection{A773}\label{subsub:A773 flux values}
The RH in A773 is bright and extended, clearly detected in both LOFAR and uGMRT images, and it is located at the center of the cluster, following the X-ray emission (Fig.~\ref{fig:A773 Low resolution}). The halo extends over a largest linear size (LLS) of $\sim$ 2 Mpc and is elongated along NE-SW direction in both LOFAR and uGMRT images aligning with the previous studies \citep{govoni01six,barrena2007internal}. 
The integrated flux density values of the halo in LOFAR and uGMRT, based on the LOFAR 2$\sigma$ contour region, are 143.4 $\pm$ 14.5 mJy and 27.1 $\pm$ 1.5 mJy, respectively. When measured using the uGMRT 2$\sigma$ contour regions, the integrated flux density values are 141.1 $\pm$ 14.3 mJy and 26.5 $\pm$  1.5 mJy in LOFAR and uGMRT respectively. 
The integrated spectral index $\alpha$ obtained in the two cases are consistent -1.10 $\pm$ 0.08. 
In applying the injection method (see Sect \ref{subsec: Lofar model injection}), 
the flux density of the injected region measured in LOFAR is 149.2 $\pm$ 15.1 mJy. Upon rescaling the model to 650 MHz with an \(\alpha\) of -1.1, the flux density obtained was 28.9 mJy. When this rescaled flux was injected into the uGMRT visibilities, a flux density of 25.6 mJy was recovered. This value is consistent within $\pm5\%$ with the flux density measured within the \(2\sigma\) contour using both LOFAR and uGMRT methods. Therefore, we can conclude that $\alpha \sim$ -1.1 is a reliable estimate of the halo spectral index, and that $\sim 28.9$ mJy is a likely value for its flux density at 650 MHz, with uGMRT experiencing a flux density loss of approximately $\sim7\%$. The integrated flux density and spectral index values using different methods are reported in Table. \ref{tab:flux and slopes}.
\subsubsection{A1351}\label{subsub:A1351 flux values}
For A1351, the process of source subtraction was challenging due to the presence of many contaminating sources that needed to be removed to isolate the diffuse radio halo emission. Since the BCG, TG, and the ridge could still leave some residuals after subtraction they were masked during the subtraction of discrete sources while the other contaminants were removed. The radio halo in A1351 is large, bright, and extended, and is clearly detected in both LOFAR and uGMRT observations. Its morphology broadly follows the distribution of the X-ray emission (Fig. \ref{fig:A1351 Low resolution}) with diffuse emission extending up to \(r_{500}\). However, the faint emission detected near and beyond \(r_{500}\) in the NE region in both the LOFAR and uGMRT images (regions A $\&$ B in Fig \ref{fig:A1351 Low resolution}) 
remains of uncertain origin. Higher sensitivity observations are necessary to discriminate the possibile contribution of discrete sources or residuals in the images. High-resolution images from LOFAR hint at the presence of a faint radio galaxy causing this emission (region A Fig \ref{fig:A1351 Low resolution}), although its location at the outskirts of the cluster raises the possibility that it could also be a relic.  This region lacks significant X-ray emission and does not show any optical counterpart in Panoramic Survey Telescope $\&$ Rapid Response System (Pan-STARRS;\citep{chambers16arx})  observation. Therefore, the size and flux density of the radio halo were measured excluding this faint emission. The halo has a LLS of $\sim$ 2.0 Mpc in both LOFAR and uGMRT. 
If the faint emission near and beyond \(r_{500}\) is included as part of the halo, the total extend of the diffuse emission increases to $\sim$ 2.6 Mpc in both the LOFAR and uGMRT images. \\
The integrated flux density values of the halo in LOFAR and uGMRT using the different methods outlined in Sect. \ref{subsec: 2 sigma contours} are reported in Table. \ref{tab:flux and slopes}. In both cases, the integrated spectral index is $-0.95 \pm 0.07$. The flux density contributed only by the ridge to the halo is 95.5 $\pm$ 9.5~mJy in LOFAR and 21.7 $\pm$ 1.1~mJy in uGMRT. When all the contaminants (BCG, tail, and ridge) are taken into account, the total contribution to the halo flux increases to 121.4 $\pm$ 12.1~mJy and 29.5 $\pm$ 1.4~mJy in LOFAR and uGMRT, respectively.
The uGMRT flux density measurement aligns (within 1$\sigma$ uncertainty) with the flux density reported by \cite{chatterjee2022unveiling}, which is $87.2 \pm 6.1 \, \text{mJy}$ (within 3 $\sigma$ contours) using GMRT at 610 MHz. 
In applying the injection method (see Sect \ref{subsec: Lofar model injection}), 
the flux density of the injected model measured by LOFAR is 408.6 $\pm$ 40.9 mJy. Rescaling the model to 650 MHz with $\alpha=-0.95$, the flux density obtained was 97.6 mJy. When this rescaled flux was injected into the uGMRT visibilities, a flux density of 88.0 mJy was recovered. This value is within $\pm5\%$ of the flux density value measured within the \(2\sigma\) contour using both LOFAR and uGMRT methods. Therefore, we can argue that $\alpha \sim -0.95$ is a reliable measurement for the radio halo spectral index and 97.6 mJy is a reliable measurement of the flux density of the halo at 650 MHz, with uGMRT experiencing a flux density loss of 4-7\%. 
\begin{table*}[h]
    \centering
    \begin{tabular}{|c|c|c|c|c|}
        \hline
        \textbf{Cluster name} & 
        \textbf{Resolution} &
        \textbf{Flux density LOFAR (mJy)} & \textbf{Flux density uGMRT (mJy)} & \textbf{Spectral index values} \\
        \hline
        \multirow{3}{*}{A773} &
        50\arcsec $\times$ 50\arcsec & 143.4 $\pm$ 14.5 & 27.1 $\pm$ 1.5 & -1.10 $\pm$ 0.08 \\
        & 50\arcsec $\times$ 50\arcsec & 141.1 $\pm$ 14.3 & 26.5 $\pm$ 1.5 & -1.10 $\pm$ 0.08 \\
        & -- & 149.2 $\pm$ 15.1 & 28.9 & -1.10 $\pm$ 0.07 \\
        \hline
        \multirow{3}{*}{A1351} &
        33\arcsec $\times$ 33\arcsec & 392.6 $\pm$ 39.3 & 93.4 $\pm$ 4.7 & -0.95 $\pm$ 0.07 \\
        & 33\arcsec $\times$ 33\arcsec & 385.2 $\pm$ 38.5 & 91.5 $\pm$ 4.6 & -0.95 $\pm$ 0.07 \\
        & -- & 408.6 $\pm$ 40.9 & 97.6 & -0.95 $\pm$ 0.07 \\
        \hline
    \end{tabular}
    \caption{Flux density and spectral index values for clusters A773 and A1351 using various methods, presented in the order of LOFAR $2\sigma$ contours, uGMRT $2\sigma$ contours and LOFAR model injection (For clarification on the methods and flux density values, see sec~\ref{subsub:A773 flux values} and~\ref{subsub:A1351 flux values}).
}
    \label{tab:flux and slopes}
\end{table*}
\subsection{Spectral index map}
\begin{figure*}[h]
    \centering
    \begin{subfigure}[b]{0.39\textwidth}
        \centering
        \includegraphics[width=\linewidth]{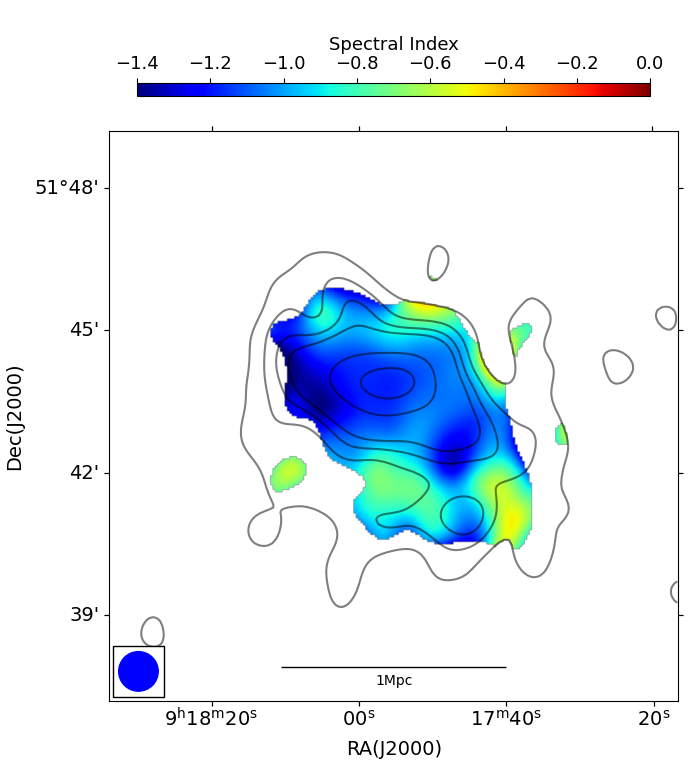}
        \caption{}
        \label{fig:spix1_con50_A773}
    \end{subfigure}
    \hspace{0.01\textwidth}
    \begin{subfigure}[b]{0.39\textwidth}
        \centering
        \includegraphics[width=\linewidth]{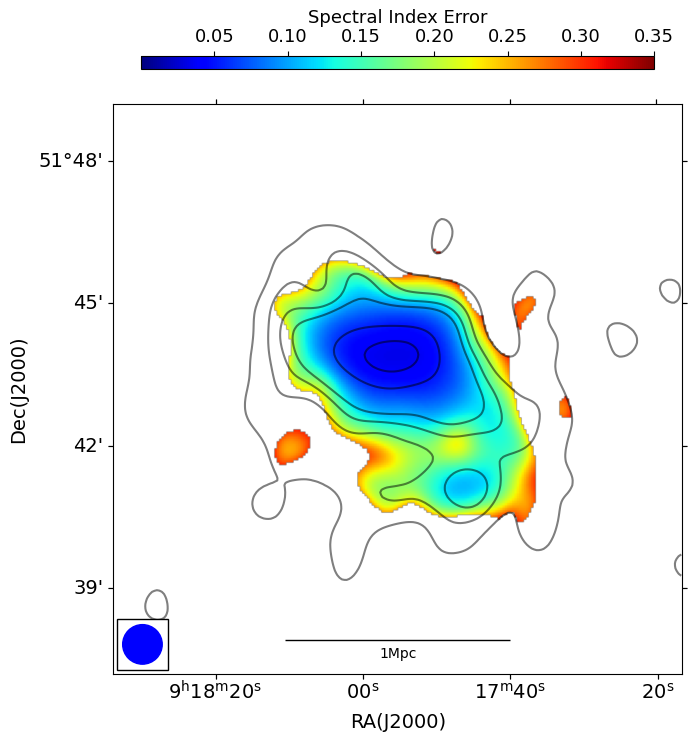}
        \caption{}
        \label{fig:spix_error_con50_A773}
    \end{subfigure}
    \caption{\textit{From left:} Spectral index map between the source subtracted images of LOFAR at 144 MHz and uGMRT at 650 MHz of A773 and its error map, with a resolution of \(50'' \times 50''\), overlaid with 3 sigma contours of the LOFAR image. The contour levels for LOFAR are set at \((3, 6, 9, 12, 27, 41) \times \sigma_{\text{rms}}\), where \(\sigma_{\text{rms}}\) is 0.30 mJy/beam.
}
    \label{fig:combined spix A773}
\end{figure*}
\begin{figure*}[h]
    \centering
    \begin{subfigure}[b]{0.39\textwidth}
        \centering
        \includegraphics[width=\linewidth]{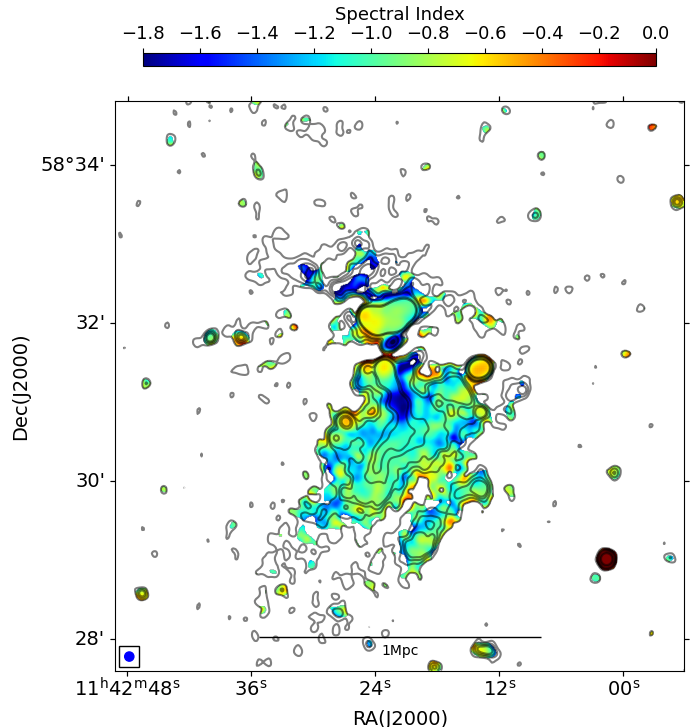}
        \caption{}
        \label{fig:spix_con7_A1351}
    \end{subfigure}
    \hspace{0.01\textwidth}
    \begin{subfigure}[b]{0.39\textwidth}
        \centering
        \includegraphics[width=\linewidth]{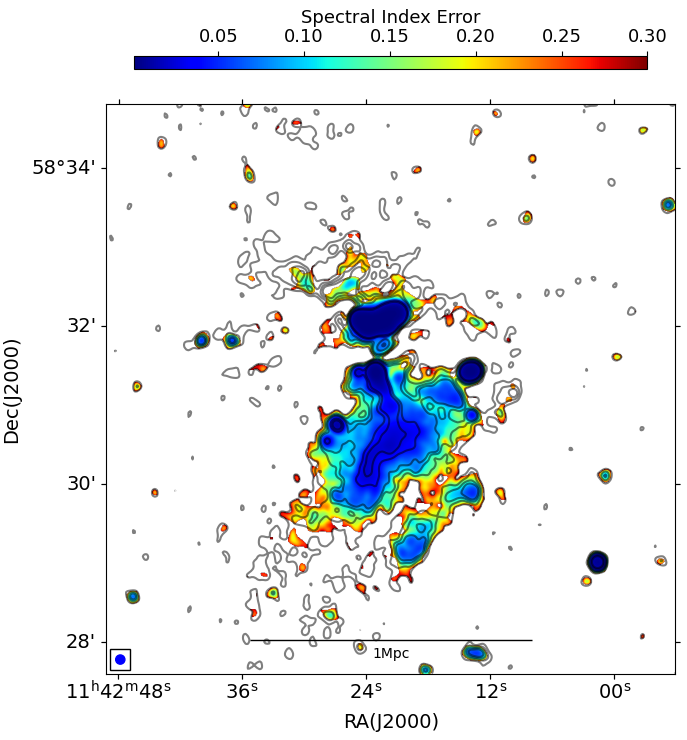}
        \caption{}
        \label{fig:spix_error_con7_A1351}
    \end{subfigure}
    \caption{\textit{From left:} Spectral index map between the high resolution images of LOFAR at 144 MHz and uGMRT at 650 MHz of A1351 and its error map, with a resolution of \(7'' \times 7''\), overlaid with 3 sigma contours of the LOFAR image. The contour levels for LOFAR are set at \((3, 6, 9, 12, 27, 41) \times \sigma_{\text{rms}}\), where \(\sigma_{\text{rms}}\) is 0.071 mJy/beam.
}
    \label{fig:combined spix HR A1351}
\end{figure*}
\begin{figure*}[h]
    \centering
    \begin{subfigure}[b]{0.39\textwidth}
        \centering
        \includegraphics[width=\linewidth]{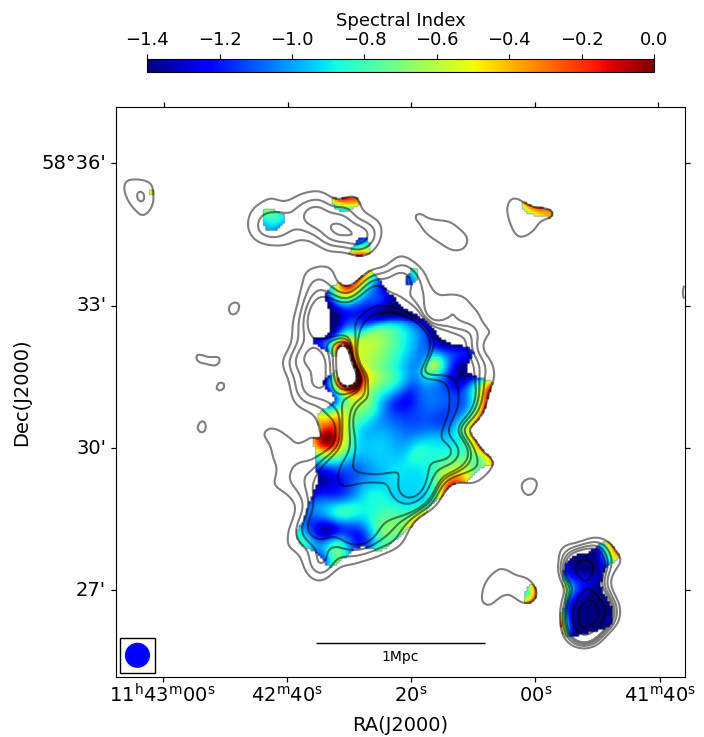}
        \label{fig:spix_con30_A1351}
    \end{subfigure}
    \hspace{0.02\textwidth}
    \begin{subfigure}[b]{0.39\textwidth}
        \centering
        \includegraphics[width=\linewidth]{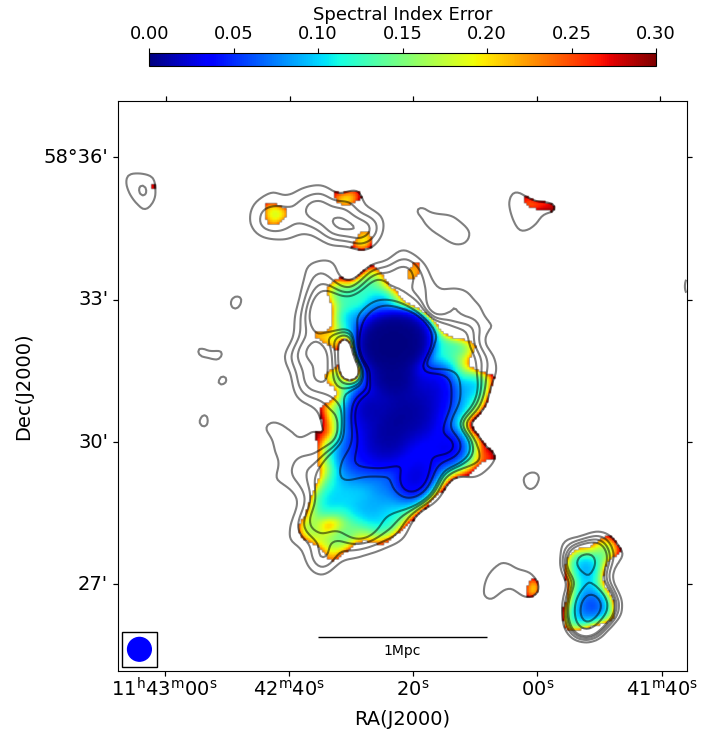}
        \label{fig:spix_error_con30_A1351}
    \end{subfigure}
    \caption{\textit{From left:} Spectral index map between the source subtracted images of LOFAR at 144 MHz and uGMRT at 650 MHz of A1351 and its error map, with a resolution of \(30'' \times 30''\), overlaid with 3 sigma contours of the LOFAR image. The contour levels for LOFAR are set at \((3, 6, 9, 12, 27, 41) \times \sigma_{\text{rms}}\), where \(\sigma_{\text{rms}}\) is 0.19 mJy/beam. \textit{Note}: The BCG, TG, and the ridge were not subtracted in the source subtracted LOFAR and uGMRT final images used to generate the spectral index map.
}
    \label{fig:combined spix LR A1351}
\end{figure*}
Spectral index (SI) maps between LOFAR 144 MHz and uGMRT 650 MHz source-subtracted images
were generated applying a common inner \textit{uv}-cut of 200$\lambda$ to sample the same spatial scales and masking regions below 3$\sigma$ threshold. 
The images were convolved to a common resolution, regridded, and aligned. Regridding was performed using \textsc{CASA}, while image alignment was carried out using \textsc{Broadband Radio Astronomy Tools (BRATS), \citep{harwood20}} Image Aligner \footnote{https://github.com/JeremyHarwood/bratsimagealignment}. The spectral index map of the aligned images was produced using \textsc{CASA} and does not account for flux density losses in uGMRT identified through the real LOFAR model injection. These losses primarily affect the faint, diffuse emission located in the peripheral regions; however, they are on the order of a few percent, as shown in Sec.~\ref{subsub:A773 flux values} and ~\ref{subsub:A1351 flux values}, and therefore do not significantly impact the spectral analysis.

\subsubsection{A773}
For A773 the resulting SI map reveals a spatial variation of the spectral index with values ranging between -1.4 and -0.6 (Fig \ref{fig:combined spix A773}). To further illustrate this variation, the spectral index measured within radio beam sized grids in the SI map is presented as a histogram (Fig.~\ref{fig:hist_A773}). In the  core of the cluster the spectral index remains constant around -1.0.
In the south-east edge of the map, the region that falls inside 6$\sigma$ contour shows spectral flattening, and this is attributed to the source subtraction residuals which becomes prominent at low resolution. This region also exhibits high spectral index error values, making the spectral index less reliable than in the other regions (see Fig \ref{fig:combined spix A773}).
\subsubsection{A1351}
For A1351, SI maps were generated for both high-resolution and low-resolution radio images. The high-resolution SI map (\(7'' \times 7''\)) is important for studying the impact of ridge and other radio features on the halo. To create the high-resolution images, we used a Briggs weighting parameter of $-1$, with no taper and an inner \textit{uv}-cut of 200$\lambda$, for both LOFAR and uGMRT datasets. The SI map was then produced following the procedure described earlier (see Fig.~\ref{fig:combined spix HR A1351}). From the high-resolution spectral index map, we observe a steepening of the spectral slope along the tail of the TG (Fig.~\ref{fig:combined spix HR A1351}). On the other hand, the ridge, does not show any significant spectral gradient; its spectral index is 
$ \sim -0.9$. A previous study by \citet{chatterjee2022unveiling}, based on an SI map between GMRT and VLA (610~MHz - 1.4~GHz), reported a spectral steepening across the ridge, suggesting the presence of a shock downstream. However, in our SI map, no clear gradient is detected across the ridge.\\
For the low-resolution, source-subtracted spectral index  map of A1351 (Fig.~\ref{fig:combined spix LR A1351}), we adopted a different approach compared to the method used for A773. Specifically, the BCG, TG, and the ridge, were not subtracted before producing the SI maps (Fig. \ref{fig:combined spix LR A1351}). The other contaminating sources were removed. This was done intentionally because they are challenging to remove and there could be some residuals of the bright tails even after subtraction. This will also allow us to investigate whether any spectral gradient exists that could suggest a connection to a shock or support the interpretation of the ridge as a radio relic. Spectral flattening is observed in the regions of TG and BCG (Fig. \ref{fig:combined spix LR A1351}) but the image does not provide significant information regarding particle acceleration processes. 
\subsection{Point-to-Point analysis}\label{subsec:ptp A773}
To investigate the connection between the thermal and non-thermal components in these galaxy clusters we performed a point-to-point analysis of their radio and X-ray surface brightness 
($I_\mathrm{Radio}$–$I_\mathrm{X}$ relation) \citep{govoni01comparison}. Several studies have conducted point-to-point analysis, revealing that clusters with radio halos generally exhibit a positive, sub-linear trend between the radio and X-ray brightness indicating narrower thermal and broader non-thermal components distribution \citep{giacintucci05,cova19,xie20,hoang21a990,santra2024deep,balboni2024chex}. 
In our paper, point-to-point analysis was conducted using the methodology described in \citet{balboni2024chex} where surface brightness was extracted from the radio ($I_R$) and X-ray ($I_X$) images by constructing a grid over the regions above the $2\sigma$ level. Each box inside the grid has an area equal to the radio beam size, ensuring that the measurements remain independent  (see Fig.~ \ref{fig:ptp boxes}).
\subsubsection{A773}
In the analysis of the A773 cluster, the source-subtracted images 
from LOFAR and uGMRT were convolved to a common resolution of \(50'' \times 50''\). Focusing on regions above the \(2\sigma\) noise level, these images were compared with the X-ray image obtained from XMM-Newton \citep{zhang2023}. 
As depicted in Figure~\ref{fig:ptp plot}, the best-fit relation between radio and X-ray surface brightness yields a slope of \(0.81 \pm 0.06\) for LOFAR-XMM and \(0.70 \pm 0.06\) for uGMRT-XMM. This sublinear correlation aligns with the expectations from the turbulent re-acceleration model of radio halos \citep{govoni01comparison,brunetti14rev,balboni2024chex}. 
The Pearson correlation coefficient \( r_p \) is \(0.87\) for LOFAR-XMM and \(0.88\) for uGMRT-XMM, with a scatter \(\sigma\) of \(0.35\) and \(0.29\) around the best-fit relation (see Table. \ref{tab:ptp_properties}).
\vspace{-0.25cm}
\subsubsection{A1351}
\begin{figure*}[htbp]
    \centering
    \setlength{\tabcolsep}{2pt}
    \begin{subfigure}[t]{0.39\textwidth}
        \centering
        \includegraphics[width=\linewidth]{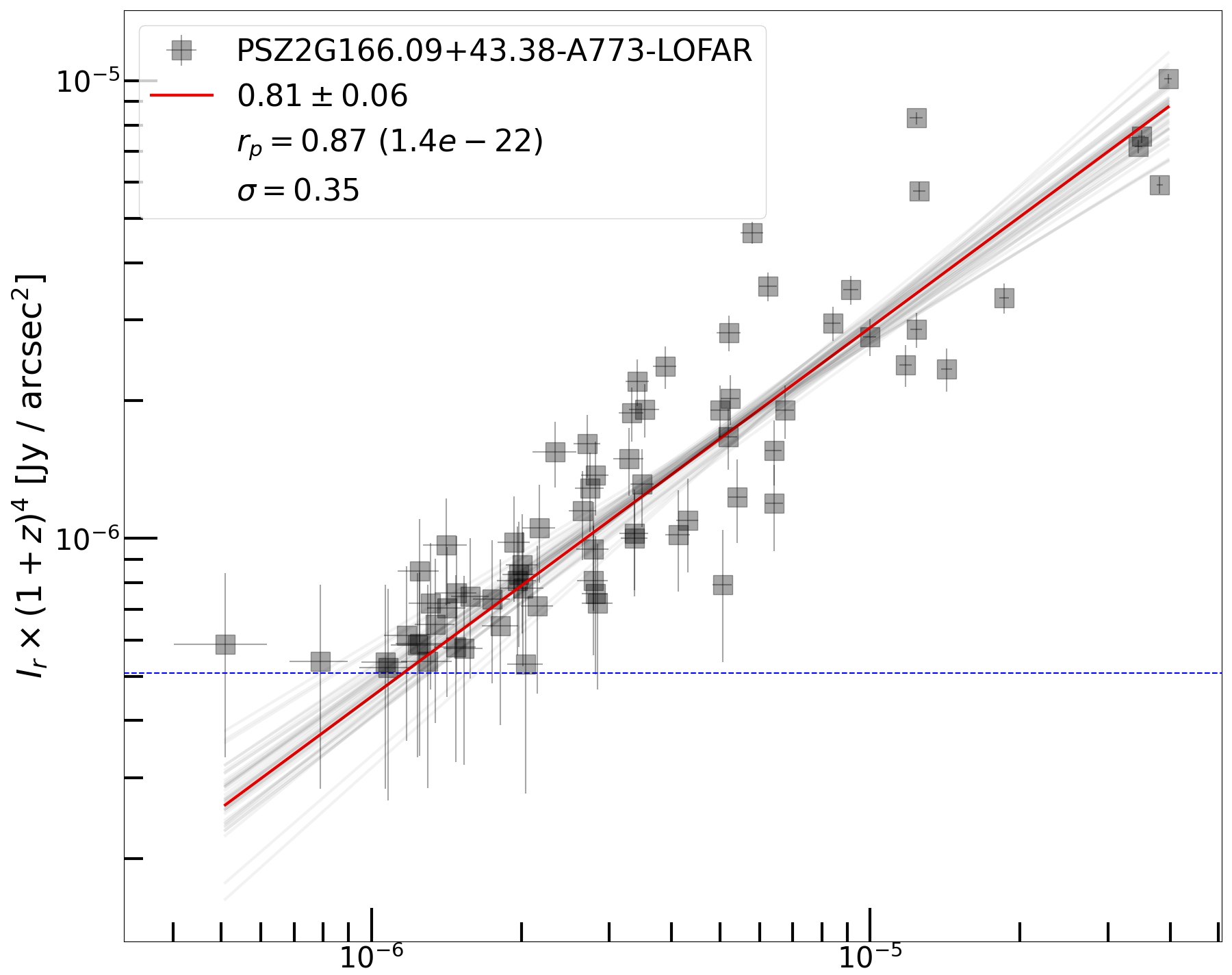}
        \label{fig:ptp plot lofar A773}
    \end{subfigure}
    \hspace{0.01\textwidth}
    \begin{subfigure}[t]{0.39\textwidth}
        \centering
        \includegraphics[width=\linewidth]{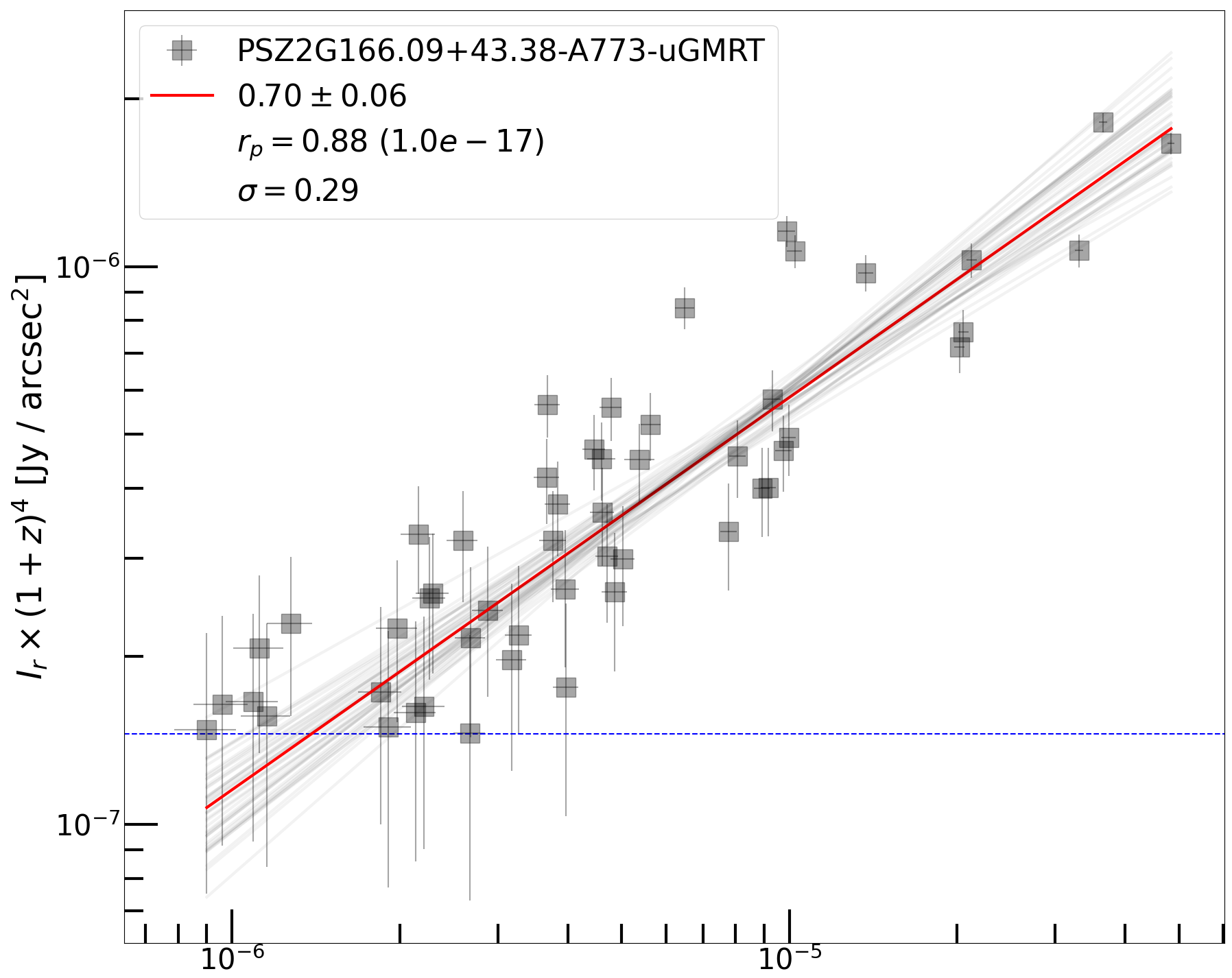}
        \label{fig:ptp plot ugmrt A773}
    \end{subfigure}

    \vspace{-2mm} 

    \begin{subfigure}[t]{0.39\textwidth}
        \centering
        \includegraphics[width=\linewidth]{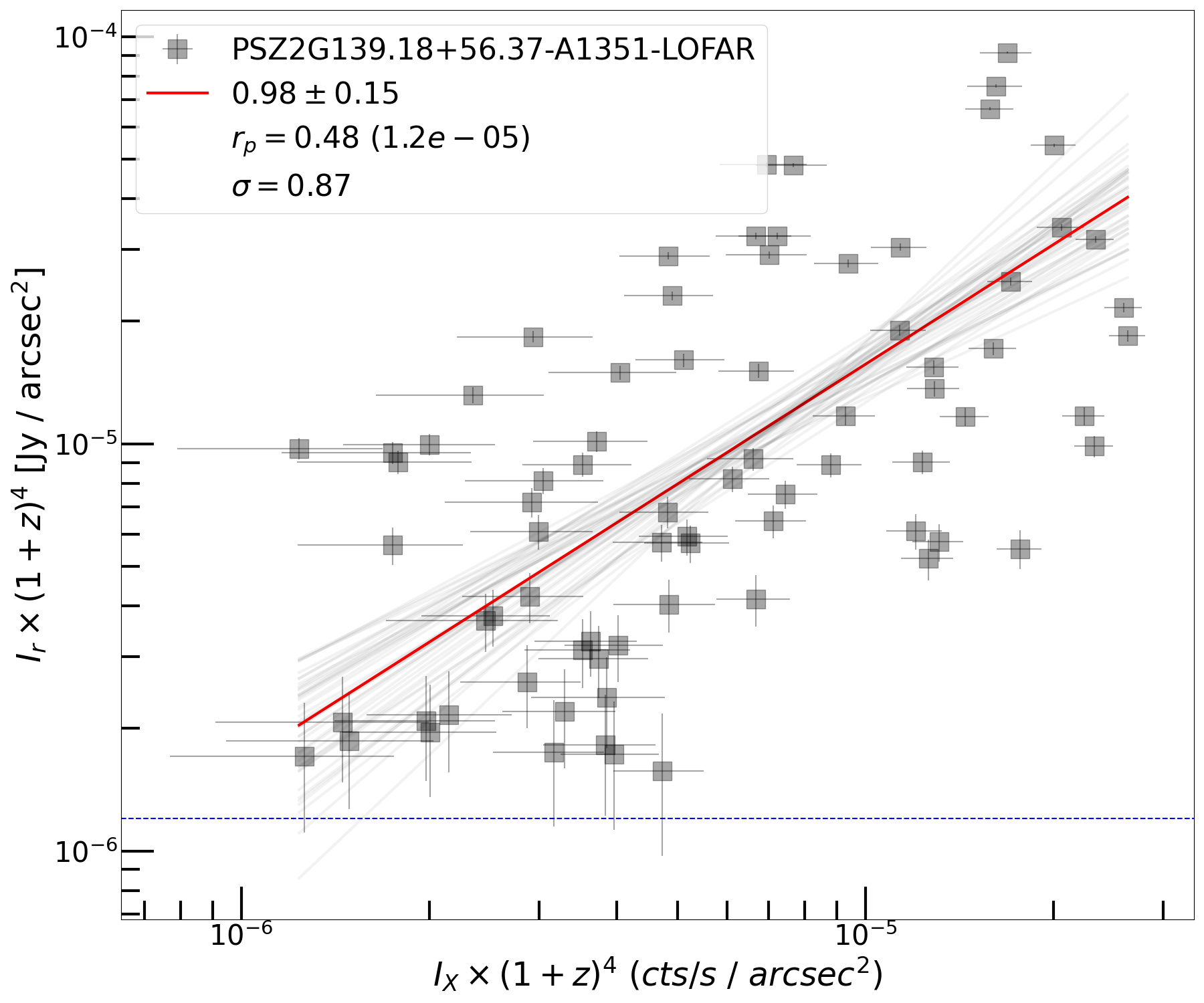}
        \label{fig:ptp plot lofar A1351}
    \end{subfigure}
    \hspace{0.01\textwidth}
    \begin{subfigure}[t]{0.39\textwidth}
        \centering
        \includegraphics[width=\linewidth]{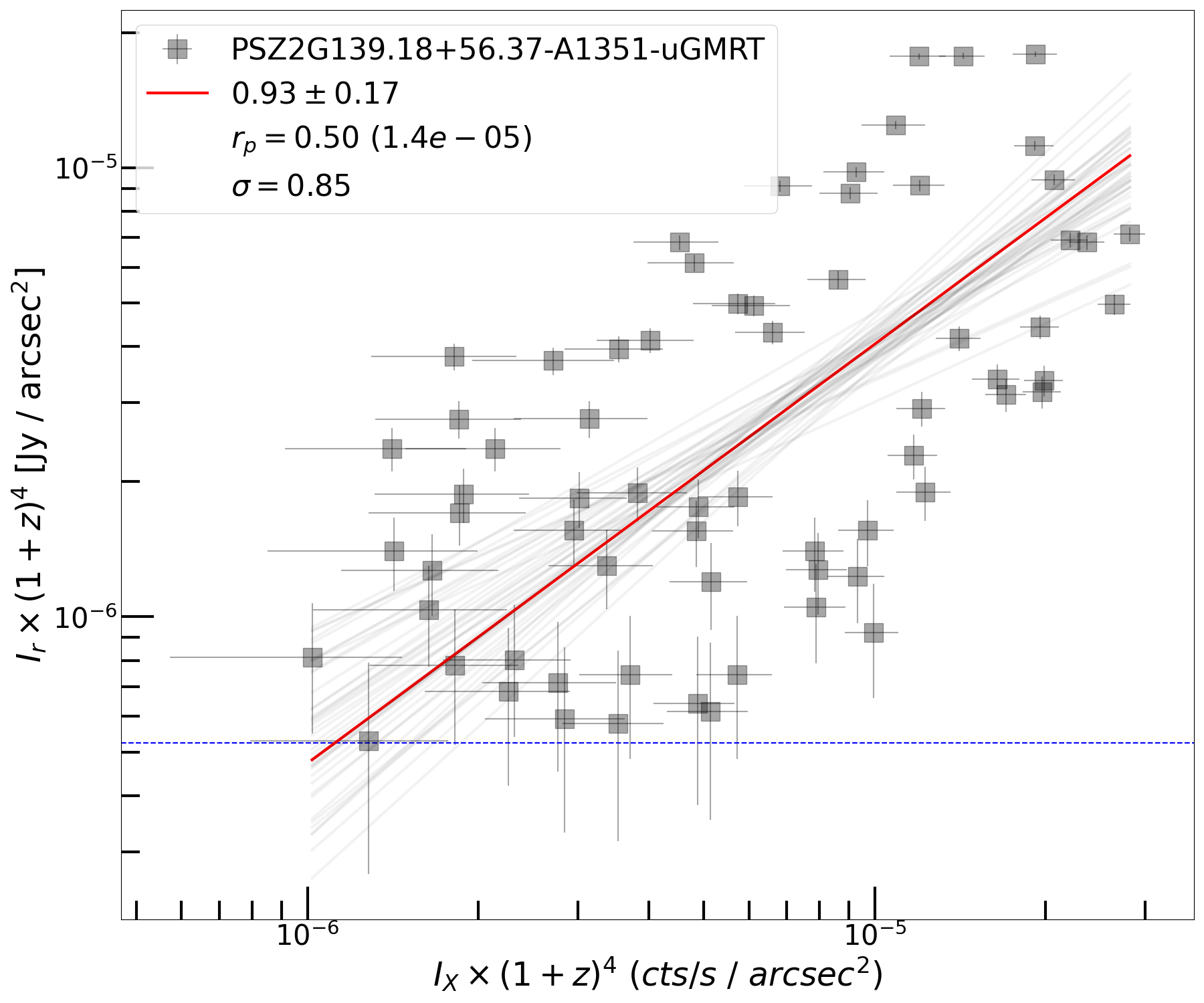}
        \label{fig:ptp plot ugmrt A1351}
    \end{subfigure}
    \vspace{-2mm}
    \caption{\textit{Top:} Radio and X-ray surface brightness comparison of LOFAR and uGMRT images at a resolution of $50'' \times 50''$ with X-ray XMM image of A773 within the regions marked by boxes in Fig.~\ref{fig:ptp boxes}. \textit{Bottom:} Same for A1351 at $33'' \times 33''$ resolution. The red line indicates the best-fit, blue dashed line is the 2$\sigma$ contour threshold. The Pearson correlation coefficient is denoted by $r_p$ with the corresponding p-values (in brackets). $\sigma$ indicates the scatter around the best-fit relation.}
    \label{fig:ptp plot}
\end{figure*}

Similar to cluster A773, the source-subtracted images for cluster A1351 
from LOFAR and uGMRT were convolved to a common resolution of \(33'' \times 33''\). Regions above the \(2\sigma\) noise level were compared with the X-ray image obtained from XMM-Newton, and their respective surface brightness within the grids (see Fig~\ref{fig:ptp boxes}) were plotted (Fig \ref{fig:ptp plot}). This analysis yielded a best-fit relation of \(0.98 \pm 0.15\) for LOFAR-XMM, with a Pearson correlation coefficient \( r_p \sim 0.48 \) and a scatter \(\sigma\) of 0.87. For uGMRT-XMM, the best-fit relation was \(0.93\pm 0.17\), with a Pearson correlation coefficient \( r_p \sim 0.50 \) and a scatter \(\sigma\) of 0.85 (see Table. \ref{tab:ptp_properties}). In both cases, the high value of the scatter might suggest a weak or absent correlation. We decided to investigate which part of the halo contributes more to the scatter, so we divided the RH into two distinct regions: the center and the outer region, fitting them separately. In both the regions, the relations were weakly correlated. A further subdivision of the RH into three spatial regions did not improve the correlation. However, this analysis showed that the scatter primarily arises from regions affected by BCG and the ridge. We did not attempt additional subdivisions due to the limited number of grids available for a reliable fit. 
\begin{table}
\caption{Point-to-point analysis of A773 and A1351. \( r_p \) denotes the Pearson correlation coefficient with the p-value in brackets. \( \sigma \) represents the scatter around the best-fit relation.}
\label{tab:ptp_properties}
\centering
\small
\hspace*{-0.5cm}
\begin{tabular}{lcc}
\toprule
\textbf{Properties} & \textbf{A773} & \textbf{A1351}\\
\midrule
Best-fit (LOFAR--XMM) & \(0.81 \pm 0.06\) & \(0.98 \pm 0.15\) \\
Best-fit (uGMRT--XMM) & \(0.70 \pm 0.06\) & \(0.93 \pm 0.17\) \\
\( r_p \) (LOFAR--XMM) & 0.87 (p=1.4e-22) & 0.48 (p=1.2e-5) \\
\( r_p \) (uGMRT--XMM) & 0.88 (p=1.0e-17) & 0.50 (p=1.4e-5) \\
\(\sigma\) (LOFAR--XMM) & 0.35 & 0.87 \\
\(\sigma\) (uGMRT--XMM) & 0.29 & 0.85 \\
\bottomrule
\end{tabular}
\end{table}
\section{Discussion}\label{sec:Discussion}
\subsection{A773}
A773 is a dynamically disturbed cluster characterized by a bimodal velocity distribution of galaxies. Optical observations from  Digitized Sky Survey (DSS) reveal two galaxy substructures with one of them aligning with the center of X-ray emission \citep{govoni04chandra}. Significant amount of substructure in A773 was first shown by ROSAT \citep{rizza98,govoni01six,starck1998structure}. 
A773 has experienced multiple merger events involving one or two massive galaxy groups. One of the major mergers was along the north–south (N–S) direction, involving the cluster and the group that contained the second dominant galaxy in the system. The most recent merger is along NEE–SWW direction, as traced by the X-ray emission \citep{barrena2007internal}. Since the X-ray peaks do not correlate with the spatial distribution of galaxies it indicates an advanced merging phase \citep{barrena2007internal}. This was further supported by the position of the dominant galaxies along the N-S direction, while the X-ray emission extending along NEE-SWW direction. \cite{barrena2007internal} also expect that the merger axis is close to the line of sight (LOS).  
\\
From the source-subtracted images of A773, it is evident that the extent of the diffuse radio emission observed in LOFAR and uGMRT is similar (Fig. \ref{fig:A773 Low resolution}). 
Contrary to the findings of \citet{govoni04chandra}, the radio halo follows the X-ray emission along NEE-SWW direction. The discrepancy may have resulted from the limited sensitivity of the previous radio and X-ray observations. 
The radio halo exhibits an integrated spectral index of               \(\alpha_{144}^{650} \sim -1.0\). The SI map (Fig.~\ref{fig:combined spix A773}) and the corresponding histogram (Fig.~\ref{fig:hist_A773}) reveal significant spatial variations in $\alpha$, suggesting the presence of complex particle acceleration processes within the cluster.

The point-to-point analysis reveals a best-fit sublinear correlation between the radio and X-ray  surface brightness distributions in both LOFAR-XMM and uGMRT-XMM dataset (Fig \ref{fig:ptp plot}). There is some scatter present in both the plots. The level of scatter is negligible and agrees with the average intrinsic scatter reported by \citet{balboni2024chex}, who investigated the radio–X-ray correlations in five other clusters hosting radio halos.
The best-fit relation for LOFAR–XMM is steeper than that for uGMRT–XMM. This difference is likely driven by the flattening of the spectral indices (see Fig.~\ref{fig:combined spix A773}) towards the halo outskirts, where the X-ray emission becomes fainter. A possible explanation for the spectral flattening is the presence of particles re-accelerated by merger-driven shocks, likely associated with the recent merger activity along the NE–SW axis. However, we cannot exclude the possibility that additional mechanisms are at work in the cluster.
\vspace{-1.25em}
\subsection{A1351} 
A1351 is also a dynamically disturbed galaxy cluster exhibiting a bimodal velocity distribution with two distinct peaks separated by approximately \(2500 \, \text{km\,s}^{-1}\) in the rest frame, indicative of an active merger 
\citep{barrena14} along the LOS. 
The X-ray peak coincides with one of the two BCGs, which is also the only one detected in radio (marked as BCG in Fig. \ref{fig:A1351_high_resolution_zoomed}). 
Analysis of Chandra X-ray data reveals four substructures along the N–S axis, two of which are associated with the BCGs and are primarily responsible for the observed bimodality in the galaxy velocity distribution. 
A1351 hosts a giant radio halo that extends along NE-SW with a LLS of $\sim$ 2 Mpc in both LOFAR and uGMRT images (Fig \ref{fig:A1351 Low resolution}) and the X-ray surface brightness distribution is elongated along N-NE to S-SW axis. The halo is asymmetrical with respect to X-ray emission, consistent with previous studies \citep{giacintucci09a1351, giovannini09, chatterjee2022unveiling, botteon22}. The presence of numerous discrete radio sources at the cluster center complicates the analysis of the radio halo, as it significantly hinders accurate source subtraction and the study of the diffuse emission.
An important feature within this galaxy cluster is the ridge, see Fig. \ref{fig:A1351_high_resolution_zoomed}. The ridge is claimed to be a radio relic in earlier research of \cite{chatterjee2022unveiling, giacintucci09a1351} and \cite{ barrena14}; but high-sensitivity and high-resolution imaging with uGMRT at 650 MHz (Figure \ref{fig:A1351_high_resolution_zoomed}) suggests an atypical morphology. Analysis of Chandra X-ray data by \cite{chatterjee2022unveiling} revealed a surface brightness discontinuity and a temperature jump, indicating a shock propagating along the SW direction.  The shock Mach number derived from radio observations and the temperature jump was found to be $\ge 1.5$. It was proposed that since the cluster merger is along the NNE-SSW axis, an axial shock could form the relic \citep{chatterjee2022unveiling}. According to \cite{barrena14}, since the ridge is situated at the external boundary of the southernmost X-ray clumps, it suggests that it could be a relic associated with a minor merger along the N-S direction. Despite previous X-ray studies favoring the interpretation of the relic, our analysis raises questions about this classification. The ridge’s morphology (Fig ~\ref{fig:A1351_high_resolution_zoomed}), lack of a spectral gradient (Figs \ref{fig:combined spix HR A1351}), and the presence of an optical counterpart in its central region (Figs.~\ref{fig:A1351_high_resolution_zoomed}) suggest that it is most likely a radio galaxy rather than a projected relic. Furthermore, the tail of the TG interacts with the head of the ridge, supplying relativistic electrons to the ridge and also to the surrounding turbulent medium, where they can be re-accelerated along with electrons from other radio sources within the cluster.

Another unique radio feature observed in A1351 is the extended emission near \( r_{500} \) (Fig \ref{fig:A1351 Low resolution}, marked as region A). This region was excluded from the calculations of the size and flux density of the radio halo because of uncertain origin. 
This region lacks any optical or X-ray counterparts. In their analysis using the VLA, \cite{giacintucci09a1351} (refer to their fig. 2) suggested the possibility of a small radio filament close to the extended emission observed near \( r_{500} \). Instead, it could potentially be associated with a radio galaxy or a relic. This is further supported by the findings of \cite{barrena14}, who dismissed the filament hypothesis due to the absence of X-ray emission or an optical subcluster in that region. Additionally, the uGMRT image at low resolution (Fig.~\ref{fig:A1351 Low resolution}, region marked B) shows a 
morphology that deviates from the LOFAR detection. Due to the morphological differences, a direct comparison between the LOFAR and uGMRT image was also not feasible.

The ridge contributes approximately one-third of the total flux density measured for the RH in LOFAR and uGMRT. The RH in A1351 has an integrated flat spectral index of \(\alpha_{144}^{650} \sim -0.95 \pm 0.07\). The low-resolution spectral index map shows no significant spectral index gradient (Fig \ref{fig:combined spix LR A1351}). Our results differ significantly from those reported by \citet{chatterjee2022unveiling} at higher frequency, who found a steeper integrated index of $\alpha = -1.72 \pm 0.33$ for the halo and $\alpha = -1.63 \pm 0.33$ for the ridge between GMRT 610 MHz and VLA 1.4 GHz. The difference in the spectral index could arise from several factors. Our analysis is based on GMRT wide band data (LOFAR--uGMRT), whereas their study used GMRT narrow band observations (GMRT--VLA), offering improved bandwidth coverage and sensitivity in our case. There are also differences in source subtraction methods, the specific regions selected for flux and spectral index measurements, and the possibility of spectral steepening at frequencies higher than 600~MHz.

A1351 is a notable cluster as it hosts a giant radio halo but deviates from the expected sublinear point-to-point correlation. The comparison between LOFAR-XMM and uGMRT-XMM surface brightness shows significant high scatter, which leads to doubts about the existence of a correlation. 
The observed correlation and scatter is likely influenced by the central contaminants which are not a part of the halo but which might hide a flatter (sub-linear) correlation. This raises the possibility that not all the observed diffuse emission is part of the halo; rather, we may be detecting only a fraction of the halo, with the rest of the emission potentially arising from other sources.\\

For both clusters, the LOFAR model injection technique was employed to determine the flux density loss of the halos in the uGMRT data-sets. This approach is a step forward from the conventional method of injecting an exponential model to find the upper limits. This method can determine the upper limits in the absence of a radio halo or quantify the flux density loss if a radio halo has been detected by the telescope. For both A773 and A1351, the flux density loss in the uGMRT data-sets ranges between 4\% and 7\%. Assessing flux density loss is crucial, as losses above 20–30\% could bias the results, producing a steeper spectral slope or underestimating the flux density.
However, in our analysis, the spectral index slopes obtained from both the methods agree within a 1$\sigma$ uncertainty. 

Accurate measurements of the spectral slope and radio flux density are essential for understanding the origin of radio halos.
In the turbulent re-acceleration model, $\alpha$ depends on cluster mass, dynamical state of the merging system, and the mass of any accreted group or cluster \citep{cassano2023planck}, providing insights into the physical mechanisms driving halo formation.

Both A773 and A1351 are are massive clusters hosting giant radio halos with integrated $\alpha_{144}^{650} \sim -1.0$. A773 shows signs of past mergers, including a recent core passage, while A1351 is undergoing an active merger. These dynamical states likely generate sufficient turbulent energy to power bright halos observable at high frequencies (e.g., 650 MHz). In addition, the presence of numerous radio galaxies around the halos may also supply seed cosmic-ray electrons for turbulent re-acceleration \citep{vazza2024seeding}.

\section{Conclusion}\label{sec:conclusion}
Our analysis on the radio halos in A773 and A1351 using LOFAR-uGMRT data allowed to derive the following results:,
\begin{itemize}
    \item The galaxy clusters A773 and A1351 are massive merging clusters both hosting giant radio halos. 
    The LLS of the radio halo in A773 and A1351 is $\sim$ 2 Mpc in both LOFAR and uGMRT images. 

    
    \item Both A773 and A1351 host numerous radio sources within the halo region. Notably, A1351 contains three significant sources that affect the halo emission: BCG, TG and a ridge. The ridge was considered a relic in \cite{chatterjee2022unveiling,giacintucci09a1351} and \cite{barrena14}. However, our high-sensitivity and high-resolution images from uGMRT at 650 MHz (Fig \ref{fig:A1351_high_resolution_zoomed}) raise a question about the nature of this emission as its morphology resembles that of a radio galaxy.
    \item The flux densities and spectral indices were derived using two approaches: measuring flux densities within 2$\sigma$ contours in LOFAR and uGMRT images, and modeling with injection of real LOFAR data. Both clusters show an integrated flat spectral index of $\sim$ -1.0, consistent with both the methods.

    \item A773 shows varying spectral index in the SI map with no clear trend. A1351 shows no significant spectral gradients in the SI map. Additionally, the ridge in A1351 did not display any spectral gradient, which is in contrast with previous studies \citep{chatterjee2022unveiling}.

    \item A point-to-point analysis 
    revealed that A773 exhibits a sublinear relation between radio and X-ray surface brightness in both LOFAR and uGMRT data, which is consistent with expectations of the turbulent re-acceleration scenario \citep{brunetti14rev}. In contrast, A1351 exhibits a weak or absent correlation between LOFAR-XMM and uGMRT-XMM, with an high scatter ($\sigma \sim$ 0.85-0.87) dominating the relationship. We suggest that the presence of numerous contaminants (discrete and diffuse sources embedded within the halo) may obscure the underlying sublinear relation.
    
    

\end{itemize}
 As a future work, conducting a LOFAR Very Long Baseline Interferometry (VLBI) and polarization analysis on A1351 could provide further insights into the nature of the ridge and accurate source subtraction.


\begin{acknowledgements}
ABon and MBal acknowledges support from the Horizon Europe ERC CoG project BELOVED, GA n. 101169773. FdG and CG acknowledge support from the ERC Consolidator Grant ULU 101086378. SC acknowledges the support from Rhodes University and the National Research Foundation (NRF), South Africa. R.C. and K.S. acknowledges financial support from the INAF grant 2023 “Testing the origin of giant radio halos with joint LOFAR-uGMRT observations” (1.05.23.05.11).
Part of The research activities described in this paper were carried out with contribution of the Next Generation EU funds within the National Recovery and Resilience Plan (PNRR), Mission 4 - Education and Research, Component 2 - From Research to Business (M4C2), Investment Line 3.1 - Strengthening and creation of Research Infrastructures, Project IR0000034 – “STILES - Strengthening the Italian Leadership in ELT and SKA”. R.K. acknowledges the support of the Department of Atomic Energy, Government of India, under project no. 12-R\&D-TFR-5.02-0700.
This paper is based on data obtained with the Giant Metrewave Radio Telescope (GMRT). We thank the staff of the GMRT that
made these observations possible. GMRT is run by
the National Centre for Radio Astrophysics of the Tata Institute of Fundamental Research. The National Radio Astronomy Observatory is a facility of the
National Science Foundation operated under cooperative agreement by Associated Universities, Inc.  LOFAR \citep{vanhaarlem13}
 is the LOw Frequency ARray designed and constructed by ASTRON. It has
 observing, data processing, and data storage facilities in several countries, which
 are owned by various parties (each with their own funding sources), and are
 collectively operated by the ILT foundation under a joint scientific policy. The
 ILT resources have benefitted from the following recent major funding sources:
 CNRS-INSU, Observatoire de Paris and Université d’Orléans, France; BMBF,
 MIWF-NRW, MPG, Germany; Science Foundation Ireland (SFI), Department
 of Business, Enterprise and Innovation (DBEI), Ireland; NWO, The Netherlands; The Science and Technology Facilities Council, UK; Ministry of Science
 and Higher Education, Poland; Istituto Nazionale di Astrofisica (INAF), Italy.
 This research made use of the Dutch national e-infrastructure with support of
 the SURF Cooperative (e-infra 180169) and the LOFAR e-infra group. The Jülich LOFAR Long Term Archive and the German LOFAR network
 are both coordinated and operated by the Jülich Supercomputing Centre (JSC),
 and computing resources on the supercomputer JUWELS at JSC were provided
 by the Gauss Centre for Supercomputing e.V. (grant CHTB00) through the John
 von NeumannInstitute for Computing (NIC). This research made use of the University of Hertfordshire high performance computing facility and the LOFAR-UK computing facility located
at the University of Hertfordshire and supported by STFC [ST/P000096/1] and Italian LOFAR-IT computing infrastructure supported and operated by
INAF, including the resources within the PLEIADI special "LOFAR" project
by USC-C of INAF, and by the Physics Department of Turin University (under the agreement with Consorzio Interuniversitario per la Fisica Spaziale) at
the C3S Supercomputing Centre, Italy. The PanSTARRS1 Surveys (PS1) and the PS1 public science archive have been made
possible through contributions by the Institute for Astronomy, the University of
Hawaii, the Pan-STARRS Project Office, the Max-Planck Society and its participating institutes, the Max Planck Institute for Astronomy, Heidelberg and the
Max Planck Institute for Extraterrestrial Physics, Garching, The Johns Hopkins
University, Durham University, the University of Edinburgh, the Queen’s University Belfast, the Harvard-Smithsonian Center for Astrophysics, the Las Cumbres
Observatory Global Telescope Network Incorporated, the National Central University of Taiwan, the Space Telescope Science Institute, the National Aeronautics and Space Administration under Grant No. NNX08AR22G issued through
the Planetary Science Division of the NASA Science Mission Directorate, the
National Science Foundation Grant No. AST–1238877, the University of Maryland, Eotvos Lorand University (ELTE), the Los Alamos National Laboratory,
and the Gordon and Betty Moore Foundation. This work is based on observations obtained with XMM-Newton, an ESA science mission with instruments
and contributions directly funded by ESA Member States and NASA.  This research has made use of SAOImageDS9, developed by Smithsonian Astrophysical Observatory \citep{joye03}.
This research made use of APLpy, an open-source plotting package
 for Python \citep{robitaille12}, Astropy, a community-developed core
 Python package for Astronomy \citep{astropy13,price2018astropy}, Matplotlib
 \citep{hunter07} and Numpy \citep{harris2020array}.
\end{acknowledgements}
\bibliographystyle{aa_url.bst}
\bibliography{Bibliography,library_additions,my_bib}
\clearpage 
\onecolumn
\begin{appendix}
\section{LOFAR model injection of radio halos}
Injection of the real LOFAR model of A773 RH into the uGMRT image at a position offset from the center without any artifacts.
\begin{figure}[htbp]
    \centering
    \begin{minipage}{0.32\textwidth}
        \centering
        \includegraphics[width=\linewidth]{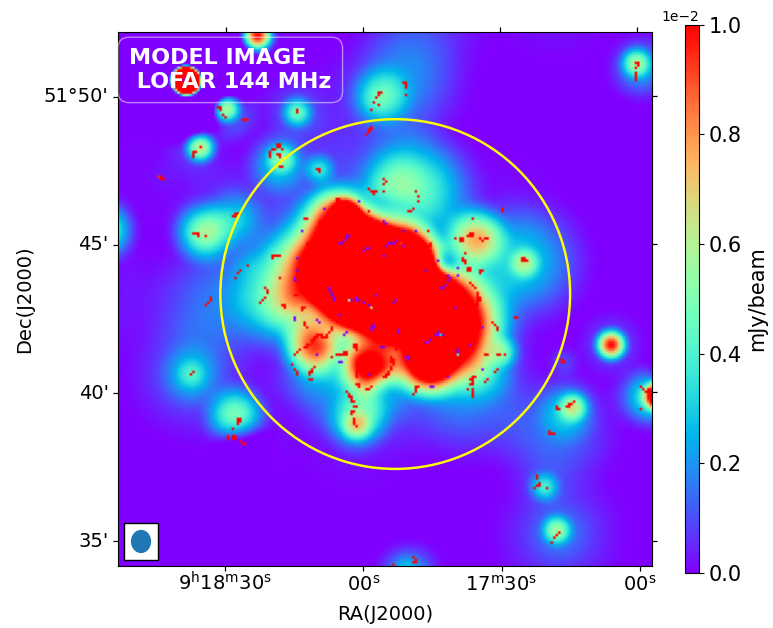}
    \end{minipage}
    \hfill
    \begin{minipage}{0.32\textwidth}
        \centering
        \includegraphics[width=\linewidth]{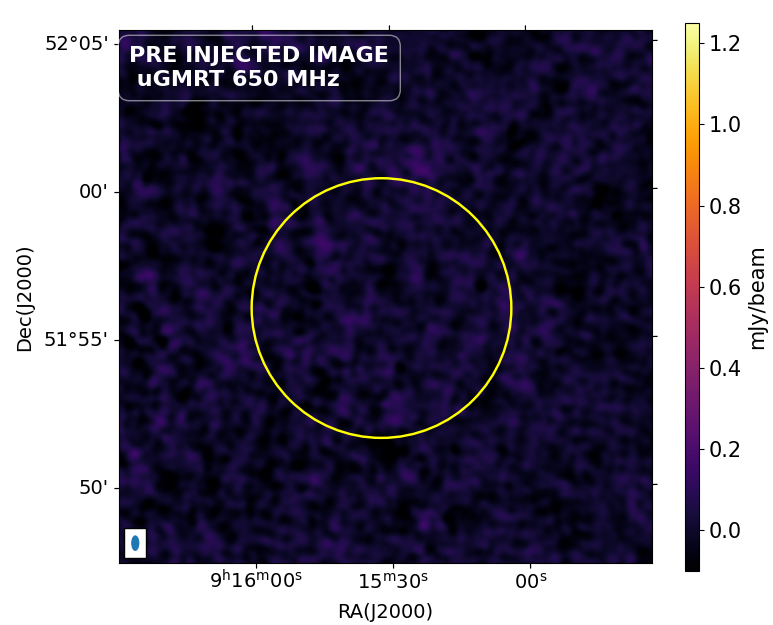}
    \end{minipage}
    \hfill
    \begin{minipage}{0.32\textwidth}
        \centering
        \includegraphics[width=\linewidth]{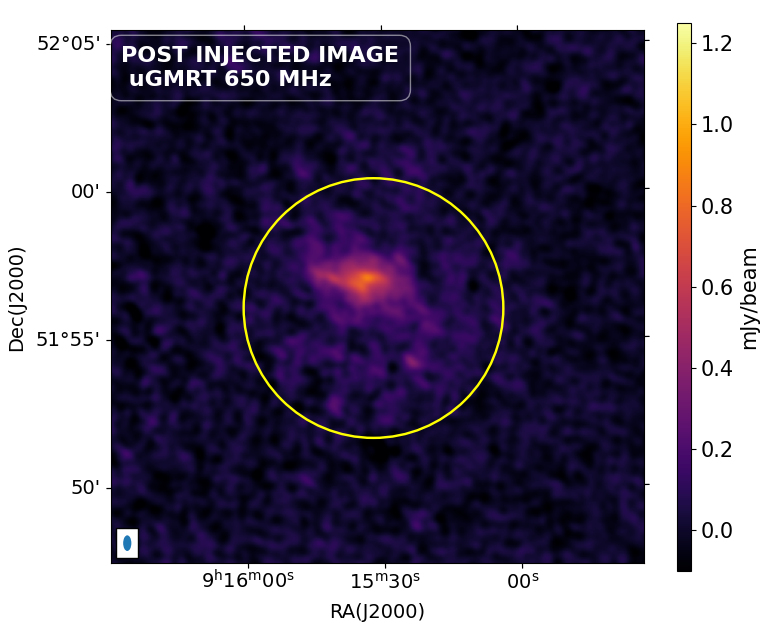}
    \end{minipage}
    \caption{Real LOFAR model injection into a uGMRT image. \textit{From left to right:} LOFAR source-subtracted model image tapered at 100 kpc (144 MHz), with the yellow region indicating the area chosen for injection; uGMRT source-subtracted image (650 MHz) before injection at the designated RA and DEC; uGMRT source-subtracted image after injection of the LOFAR model rescaled at 650 MHz with $\alpha=-1$, with the yellow region representing the size of the halo recovered after the injection.}
    \label{fig:injected_images_sec2}
\end{figure}

\section{Distribution of \(\alpha\) in the SI map}
Histogram showing the distribution of spectral index ($\alpha$) values from the SI map in Fig.~\ref{fig:combined spix A773}. The $\alpha$ values were averaged within radio beam sized boxes across the map and plotted for regions above the 3$\sigma$ and 5$\sigma$ contours.

\begin{figure*}[h]
    \centering
    \includegraphics[width=0.5\textwidth]{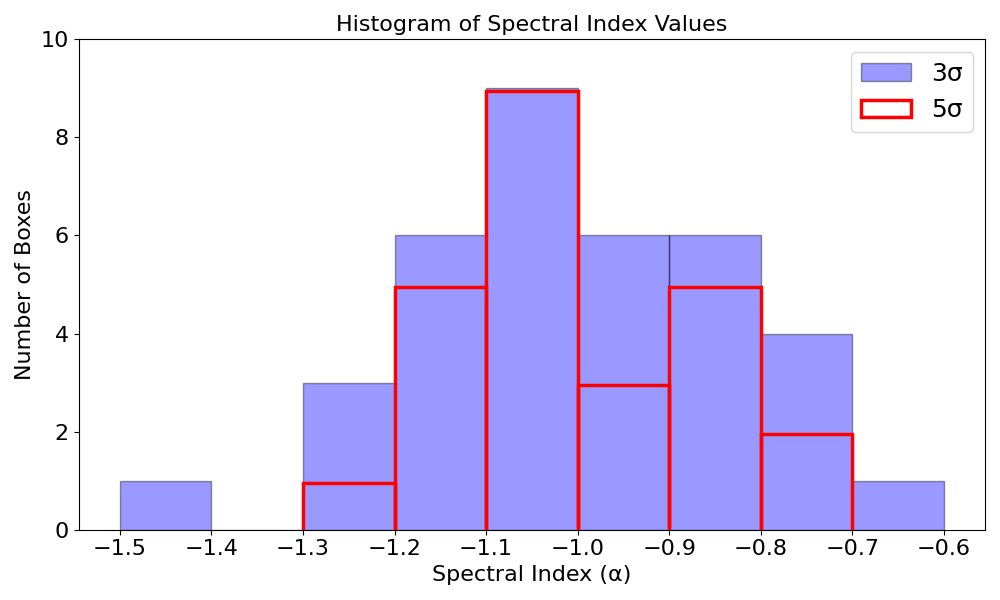}
   \caption{Distribution of spectral index ($\alpha$) values in A773, measured within radio beam sized boxes ($50\arcsec \times 50\arcsec$). Blue and red bins represent regions with flux densities above $3\sigma_{rms}$ and $5\sigma_{rms}$, respectively.}

   \label{fig:hist_A773}
\end{figure*}

\vspace{-1.5em}
\section{Point to point analysis}
The region of interest for the point-to-point analysis was divided into boxes with areas equivalent to the beam size, ensuring that the measurements are independent. The radio ($I_{\mathrm{radio}}$) and X-ray surface brightness ($I_{\mathrm{X}}$) were extracted for these grid cells and subsequently used for the point-to-point correlation analysis (see Fig~\ref{fig:ptp plot}). Shown here are the sampling grids used for A773 and A1351, overlaid on both the LOFAR and uGMRT images. 

\begin{figure*}[htbp]
    \centering
    \begin{subfigure}{0.45\textwidth}
        \centering
        \includegraphics[width=\linewidth]{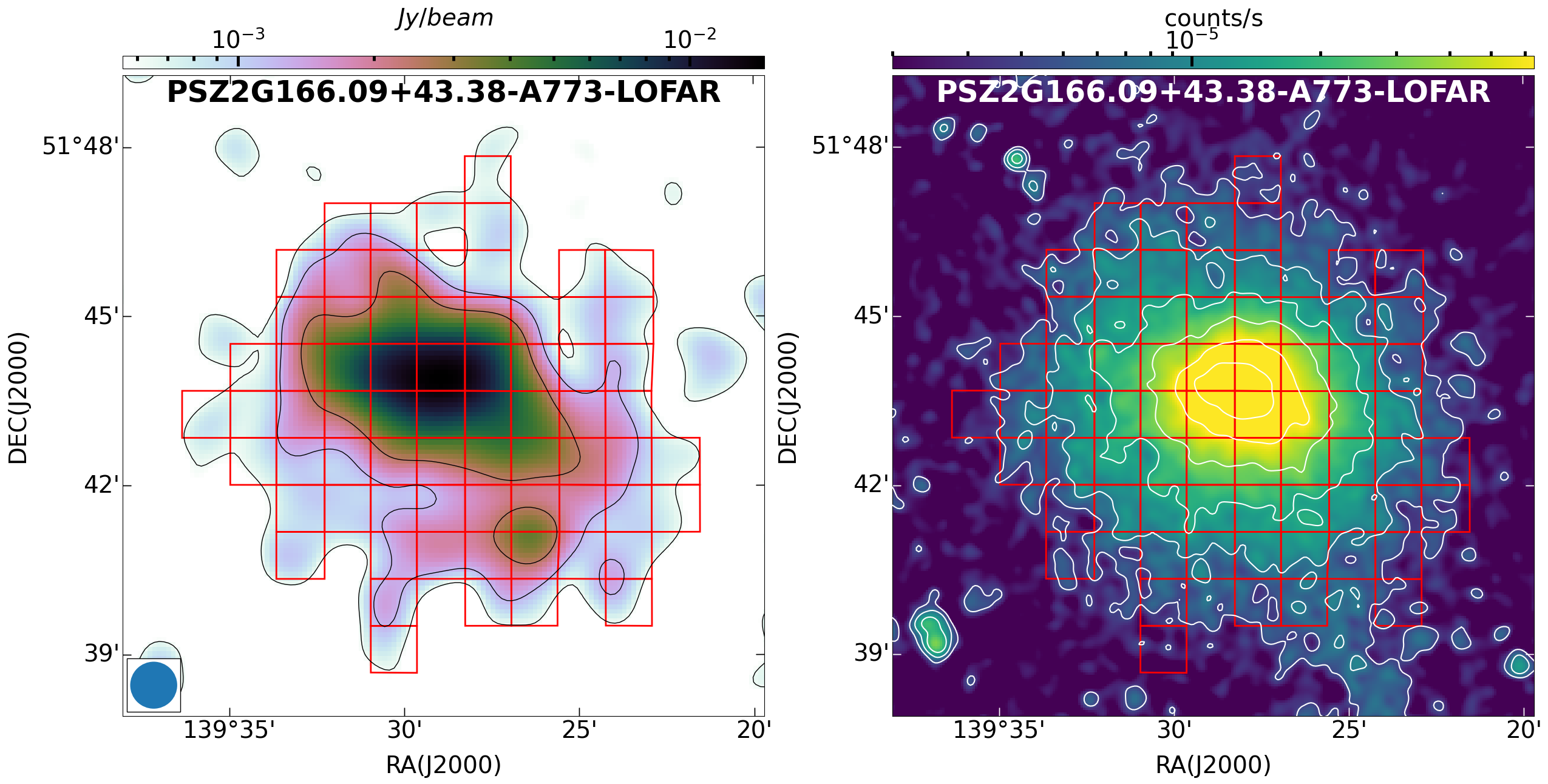}
        \label{fig:ptp boxes A773 lofar}
    \end{subfigure}
    \hspace{0.01\textwidth}
    \begin{subfigure}{0.45\textwidth}
        \centering
        \includegraphics[width=\linewidth]{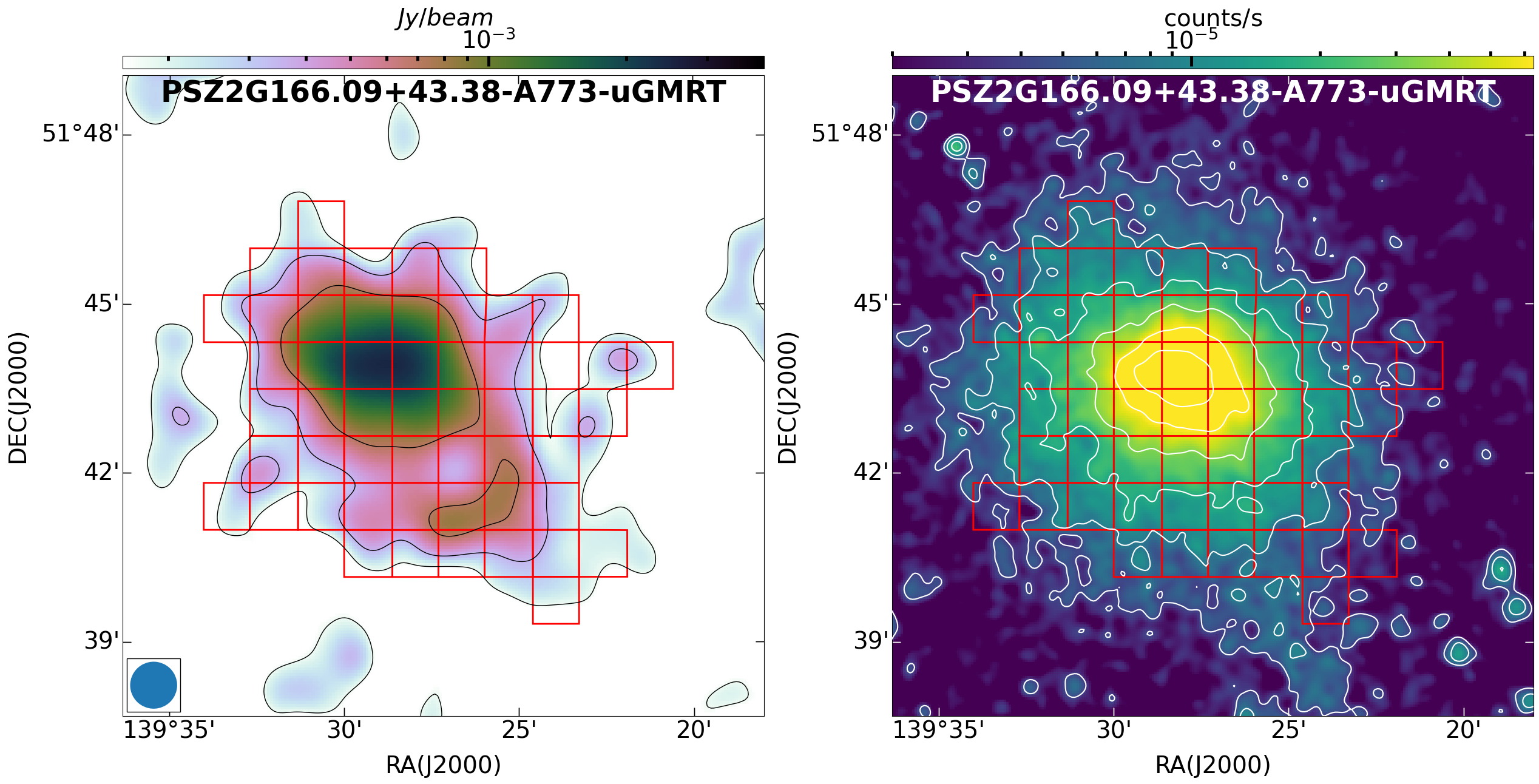}
        \label{fig:ptp boxes A773 ugmrt}
    \end{subfigure}

    \begin{subfigure}{0.45\textwidth}
        \centering
        \includegraphics[width=\linewidth]{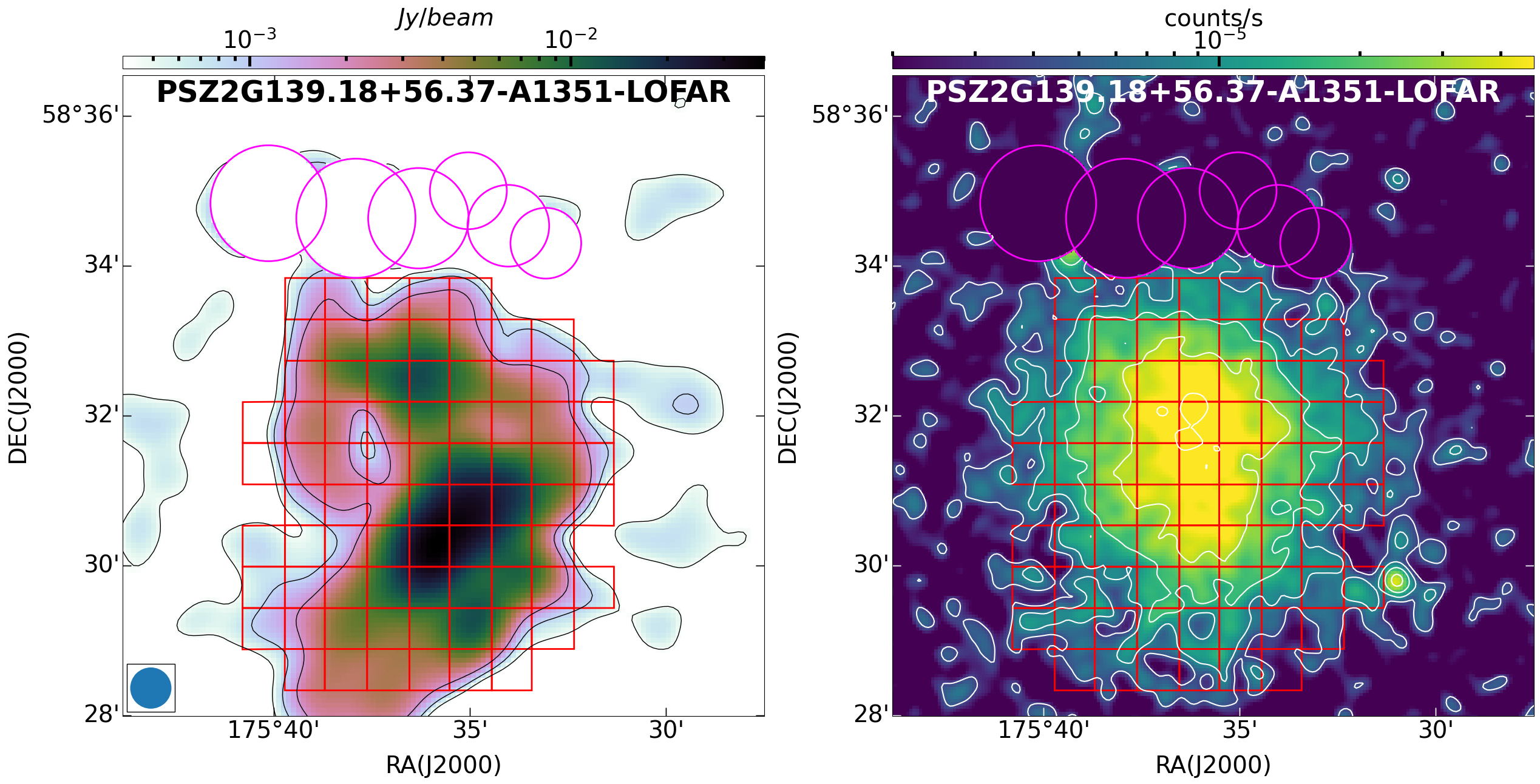}
        \label{fig:ptp boxes A1351 lofar}
    \end{subfigure}
    \hspace{0.01\textwidth}
    \begin{subfigure}{0.45\textwidth}
        \centering
        \includegraphics[width=\linewidth]{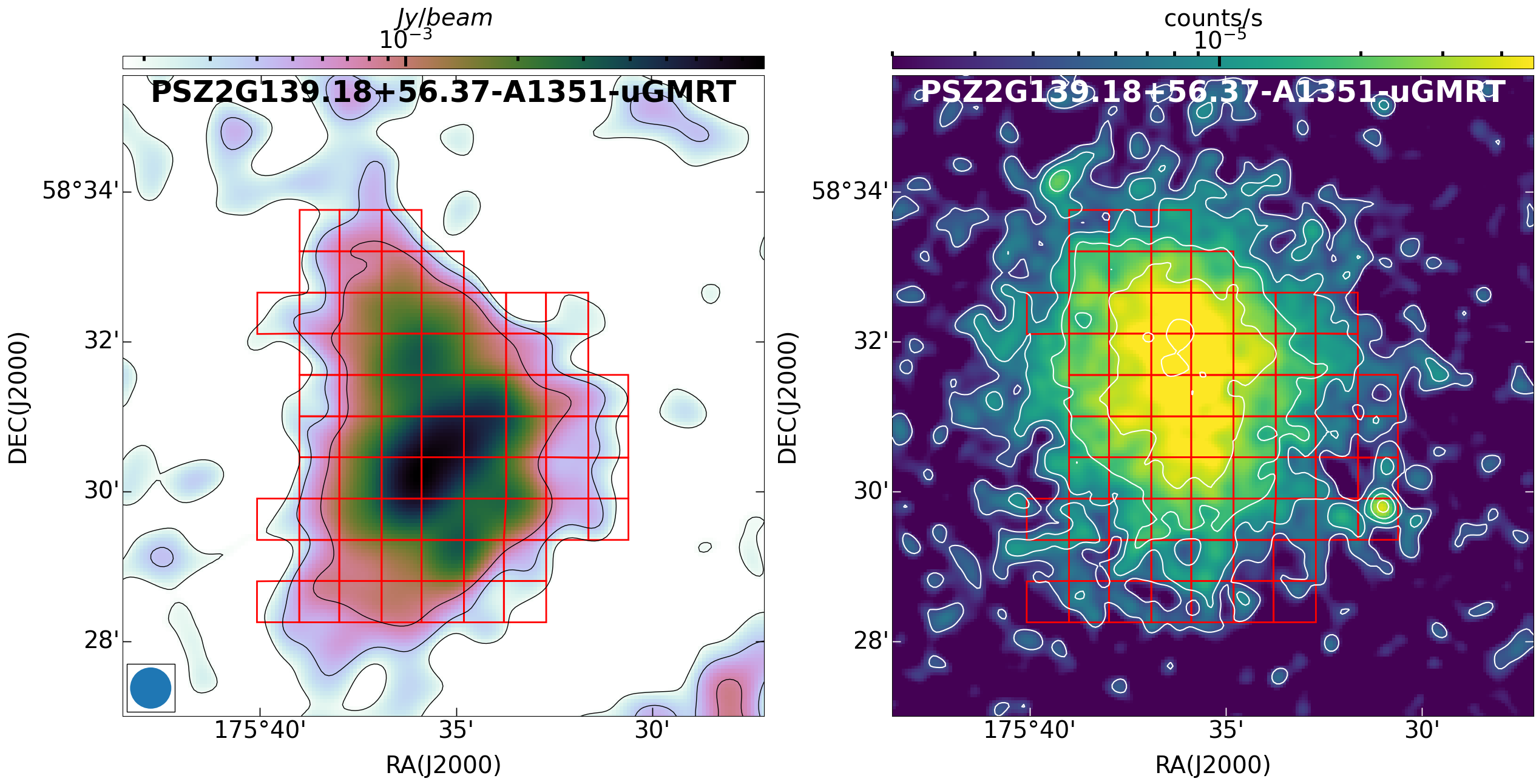}
        \label{fig:ptp boxes A1351 uGMRT}
    \end{subfigure}
    \caption{\textit{Left to right:} LOFAR source subtracted image of A773 at a resolution of $50'' \times 50''$ compared with X-ray XMM image, uGMRT source subtracted image of A773 at a resolution of $50'' \times 50''$ compared with X-ray XMM image, LOFAR source subtracted image of A1351 at a resolution of $33'' \times 33''$ compared with X-ray XMM image, uGMRT source subtracted image of A1351 at a resolution of $33'' \times 33''$ compared with X-ray XMM image. The boxes indicate the corresponding regions where radio and X-ray brightness were compared, and the circles indicate the masked regions in radio and X-ray images.}
    \label{fig:ptp boxes}
\end{figure*}

\end{appendix}
\end{document}